\documentclass[prx,twocolumn,english,superscriptaddress,floatfix,longbibliography]{revtex4-2} 

\usepackage{enumerate,amsmath}
\usepackage{hyperref,graphicx}
\usepackage{color}
\graphicspath{{./Figures}}
\usepackage{here}
\usepackage{xcolor}
\hypersetup{
    colorlinks=true,
    citecolor=blue,
    linkcolor=magenta,
    urlcolor=magenta,
}

\usepackage{braket}
\usepackage{comment}
\usepackage{algorithmic}
\usepackage{algorithm}

\newcommand{\mO}{\mathcal{O}}
\newcommand{\mP}{\mathcal{P}}
\newcommand{\mS}{\mathcal{S}}

\newcommand{\poly}{{\rm poly}}

\newcommand{\Abs}[1]{\left| #1 \right|}

\newtheorem{problem}{Problem}
\newtheorem{definition}{Definition}

\makeatletter
\renewcommand*{\@opargbegintheorem}[3]{\trivlist
      \item[\hskip \labelsep{\bfseries #1\ #2}] \textbf{(#3)}\ \itshape}
\makeatother

\begin{document}

\title{Quantum Circuit Unoptimization}

\author{Yusei Mori}
\email{u839526i@ecs.osaka-u.ac.jp}
\affiliation{Graduate School of Engineering Science, Osaka University, 1-3 Machikaneyama, Toyonaka, Osaka 560-8531, Japan}

\author{Hideaki Hakoshima}
\affiliation{Graduate School of Engineering Science, Osaka University, 1-3 Machikaneyama, Toyonaka, Osaka 560-8531, Japan}
\affiliation{Center for Quantum Information and Quantum Biology, Osaka University, 1-2 Machikaneyama, Toyonaka, Osaka 560-0043, Japan}

\author{Kyohei Sudo}
\affiliation{Graduate School of Engineering Science, Osaka University, 1-3 Machikaneyama, Toyonaka, Osaka 560-8531, Japan}

\author{Toshio Mori}
\affiliation{Center for Quantum Information and Quantum Biology, Osaka University, 1-2 Machikaneyama, Toyonaka, Osaka 560-0043, Japan}

\author{Kosuke Mitarai}
\affiliation{Graduate School of Engineering Science, Osaka University, 1-3 Machikaneyama, Toyonaka, Osaka 560-8531, Japan}
\affiliation{Center for Quantum Information and Quantum Biology, Osaka University, 1-2 Machikaneyama, Toyonaka, Osaka 560-0043, Japan}

\author{Keisuke Fujii}
\email{fujii.keisuke.es@osaka-u.ac.jp}
\affiliation{Graduate School of Engineering Science, Osaka University, 1-3 Machikaneyama, Toyonaka, Osaka 560-8531, Japan}
\affiliation{Center for Quantum Information and Quantum Biology, Osaka University, 1-2 Machikaneyama, Toyonaka, Osaka 560-0043, Japan}
\affiliation{Center for Quantum Computing, RIKEN,  2-1 Hirosawa, Wako, Saitama 351-0198, Japan}

\begin{abstract}
Optimization of circuits is an essential task for both quantum and classical computers to improve their efficiency.
In contrast, classical logic optimization is known to be difficult, and a lot of heuristic approaches have been developed so far. 
In this study, we define and construct a quantum algorithmic primitive called quantum circuit unoptimization,
which makes a given quantum circuit complex by introducing some redundancies while preserving circuit equivalence, i.e., the inverse operation of circuit optimization.
Using quantum circuit unoptimization, we propose the quantum circuit equivalence test, a decision problem contained both in the NP and BQP classes but is not trivially included in the P class. 
Furthermore, as a practical application, we construct concrete unoptimization recipes to generate compiler benchmarks and evaluate circuit optimization performance using Qiskit and Pytket.
Our numerical simulations demonstrate that quantum circuit unoptimizer systematically generates redundant circuits that are challenging for compilers to optimize, which can be used to compare the performance of different compilers and improve them.
We also offer potential applications of quantum circuit unoptimization, such as generating quantum advantageous machine learning datasets and quantum computer fidelity benchmarks. 
\end{abstract}

\maketitle

\section{Introduction}
The logic optimization problem~\cite{micheli1994synthesis} is a problem in computer science and complexity theory that aims 
to minimize the size or depth of a Boolean circuit, a mathematical model for digital logic circuits. 
This problem is essential because smaller and shallower circuits can be more efficiently implemented and executed on hardware, leading to faster and more energy-efficient computation.
The circuit optimization problem has been recognized as a challenging computational problem since the 1970s when Cook and Levin defined the concept of NP-completeness~\cite{cook1971complexity,levin1973universal,trakhtenbrot1984survey}.
Specifically, the circuit minimization problem for multi-output Boolean functions belongs to NP-hard~\cite{ilango2020np}, but whether this problem is NP-complete for single-output Boolean functions remains an unresolved question.

From a practical standpoint, circuit optimization in quantum computers has also become crucial in recent years.
The power of noisy intermediate-scale quantum (NISQ) computers~\cite{preskill2018quantum} is greatly limited both by the number of qubits and by the noise level of the gates~\cite{brandhofer2021special}.
Therefore, developing efficient software that can simplify quantum circuits is critical to fully exploiting the potential ability of NISQ.
Quantum compilers~\cite{10.1145/3519939.3523433,kharkov2022arline} are software tools designed to translate a high-level description of a quantum algorithm into a low-level, hardware-specific instruction that can be executed on a real quantum computer. 
In addition to transforming an input circuit, modern quantum compilers engage in circuit optimization, including gate reduction and gate rearrangement, aiming to achieve faster execution and mitigate gate errors in a quantum circuit.
The optimization process requires unitary equivalence between input and output circuits, and recent studies have shown such general circuit transformation rules~\cite{Cl_ment_2023, minimum_eq_theory}.

In this study, we initiate a study of a quantum algorithmic primitive called {\it quantum circuit unoptimization},
a reverse operation of the optimization of quantum circuits. 
Quantum circuit unoptimization maps the original circuit to another with redundancy while preserving functionality. 
Although this operation may seem useless, 
it offers fundamental and practical applications. 
As a theoretical application, we delve into quantum circuit equivalence tests.
When using an unoptimizer for quantum circuits, we can define an NP problem, where circuit equivalence can be verified by providing an unoptimization recipe as proof. Moreover, under a natural promise, we provide a BQP problem for a quantum circuit equivalence test, making it a decision problem that belongs to an intersection of NP and BQP problems and is not obvious to be in P.
Since the solution to this problem only requires preparing the all-zeros state, applying quantum gates, and sampling to check the probability of obtaining the all-zeros outcome, it could be efficiently implemented on NISQ devices.

In the aspect of practical applications, we aim to use the unoptimizer to generate benchmark circuits for quantum compilers such as Qiskit~\cite{Qiskit} and Pytket~\cite{Pytket}. The recipe for unoptimization includes four types of elementary gate operations: insertion, swapping, decomposition, and synthesis, which create a redundant quantum circuit preserving its equivalence. A compiler then attempts to minimize a circuit volume, and we obtain the optimized circuit depth. We can quantitatively assess compiler performance by analyzing the shift before and after compilation.

Additionally, we introduce two recipe generation options: {\it random} and {\it concatenated}. They differ in terms of whether the positions of operations depend on a previous operation or not. As a result, these two approaches represent different circuit compression performances between quantum compilers attributed to the modules included in them. Therefore, our proposed recipe for unoptimization can be used to compare different compilers and enhance their performance.

The rest of the paper is organized as follows. In Sec.~\ref{sec: def_unopt}, we define quantum circuit unoptimization, propose the equivalence tests through this operation, and describe their computational complexity.
Next, we explain the basic structure of unoptimization recipes and the procedure of quantum compiler benchmark in Sec.~\ref{sec: benchmark}.
Also, we show the detailed results of the numerical experiments.
Finally, we present a conclusion and discuss potential future directions of quantum circuit unoptimization in Sec.~\ref{sec: conclusion_and_discussion}.

\section{Definition of Quantum Circuit Unoptimization}
\label{sec: def_unopt}

In this section, we describe the definition of quantum circuit unoptimization and bring up some decision problems using this technique.
Then, we consider their computational complexity for classical and quantum computers.

Suppose $U$ is an $n$-qubit quantum circuit consisting 
of a sequence of $\poly(n)$ elementary unitary gates, 
i.e., $U=\prod _i^{\poly(n)} U_i$.
Quantum circuit unoptimization is defined from 
an original quantum circuit $U= \prod _i^{\poly(n)} U_i$ and 
unoptimization recipe $C$ composed of $\poly(n)$ classical bits. This recipe provides the information to generate another quantum circuit $V=\prod_j^{\poly(n)} V_j$, where $U$ and $V$ are equivalent.

\begin{definition}[Quantum circuit unoptimization]
\label{definition_of_unopt}
 Let $U=\prod_i^{\poly(n)} U_i$ a quantum circuit consisting of unitary gates $\{ U_i\}$. Quantum circuit unoptimization is defined by unoptimizer $\mO$ to generate
\begin{equation}
    V = \mO(U,C)
    \label{eq:def_unopt}
\end{equation}
where $V= \prod _j^{\poly(n)} V_j$ is another quantum circuit whose action is equivalent to $U$ (i.e., these two circuits satisfy $UV^{\dagger} = I$ up to global phase) and $C$ is a recipe to generate $V$. The recipe is assumed to be reversible, meaning we can put $V$ back into the original circuit $U$ by $C$. 
The recipe $C$ in Eq.~(\ref{eq:def_unopt}) is constructed from a finite set of elementary unoptimization operations $\mS$. We also assume obtaining the description of $V=\prod _j^{\poly(n)} V_j$ from $U= \prod _i^{\poly(n)} U_i$ is efficiently done on a classical computer.
\end{definition}

An example of elementary unoptimization operation is inserting redundant gates $W$ and $W^{\dagger}$, merging and decomposing unitary matrices, and so on.
Any operation that can be described using $\poly(n)$ classical bits and that preserves the quantum circuit identity can be employed as an element of unoptimization operation set $\mS$ to construct a recipe $C$.
One possible choice is to define the set $\mS$ that satisfies completeness in the sense that any two equivalent quantum circuits can be transformed into one another using the set $\mS$, known as complete equational theory for quantum circuits~\cite{Cl_ment_2023}. For instance, this set can be defined by Euler decompositions and the insertion of multi-controlled identity gates~\cite{minimum_eq_theory}. In this study, for simplicity,
we show a concrete example of the unoptimization recipe in Sec.~\ref{sec: benchmark}, which can generate an unoptimized quantum circuit that is thought to be challenging to optimize, at least on current existing quantum compilers.

Then, we pick up some types of quantum circuit equivalence tests.
First, we consider the problem of determining the equivalence of two circuits by checking the existence of the recipe $C$ that transforms one circuit into the other.
From Definition~\ref{definition_of_unopt}, once $U$ and $C$ are provided,
we can obtain the description of $V=\prod _i^{\poly(n)} V_i = \mO(U,C)$.
In Problem~\ref{equivalece_test_under_S}, we define an NP problem about the equivalence of two different quantum circuits, where $C$ works as proof.

\begin{problem}[Equivalence test under a finite unoptimization operation set] 
\label{equivalece_test_under_S}
Given a pair of quantum circuits $\{ U, V\}$ and a specific unoptimization operation set $\mS$. The problem is to decide whether two quantum circuits, $U$ and $V$, are equivalent within the finite unoptimization operation set $\mS$. If a recipe $C$ exists that satisfies $V= \mO(U, C)$, such an instance belongs to YES. For any recipe $C$, if $V\neq \mO(U, C)$, such an instance belongs to NO.
\end{problem}

The above quantum circuit equivalence test belongs to the NP class since if the recipe $C$ is given as proof, we can efficiently verify that $V=\mO(U, C)$. 
Note that the choice of the unoptimization operation set to be used as input significantly influences the difficulty of Problem~\ref{equivalece_test_under_S}. For instance, if the only allowed unoptimization operation is inserting redundant single-qubit gates into a quantum circuit, checking the existence of a recipe for converting $U$ to $V$ would be simple. However, by combining complex unoptimization operations across multiple qubits, we expect to construct a recipe whose problem complexity is so difficult that including the problem in the P class is nontrivial.

Second, consider the problem of determining the equivalence of two circuits by checking the inner product.
This type of equivalence test can be regarded as a generalization of the non-identity check and is known to be a QMA-complete problem~\cite{qma_complete}.
However, in this study, we simplify the problem of determining equivalence by checking the quantum states obtained when each quantum circuit acts on the all-zeros state, i.e., measuring the fidelity
\begin{equation}
    F(U\ket{0^n},V\ket{0^n}) = \Abs{\braket{0^n|V^{\dagger}U|0^n}}^2.
    \label{eq: fidelity}
\end{equation}
This value is obtained by acting $V^{\dagger}U$ on the all-zeros state $\ket{0^n}$ and measuring all qubits on the computational basis~\cite{pure_pure_fidelity}.

\begin{problem}[Equivalence test using inner product] 
\label{inner_product_equivalence_test}
Given a pair of quantum circuits $\{ U, V\}$.
The problem is defined to decide whether two quantum circuits, $U$ and $V$, are equivalent or not under an action on the all-zeros state; 
if $F(U\ket{0^n},V\ket{0^n}) \geq 1-1/\poly(n)$, the instance belongs to YES. Otherwise $F(U\ket{0^n},V\ket{0^n}) \leq 1-2/\poly(n)$, the instance belongs to NO. 
\end{problem}

This problem is known to be BQP-complete~\cite{pure_pure_fidelity}, indicating that it can be efficiently solved by a quantum computer but is considered difficult for a classical computer.
It is important to note that this test assumes both quantum circuits are unitary. In practical situations, especially with NISQ devices, noise, such as amplitude damping, can affect the results. Consequently, while this test is efficiently solvable on an ideal, noiseless quantum computer, its practical implementation on NISQ devices requires careful noise management and error mitigation techniques~\cite{mitigation}.

Finally, we combine these two tests to define Problem~\ref{Promised_equivalence_test}, which can be efficiently solved using a quantum computer but is challenging for a classical computer to solve. Note that the quantum circuit $V$ given as input contains a promise.
\begin{problem}[Promised version] 
\label{Promised_equivalence_test}
The input of the problem is a pair of quantum circuits $\{ U, V\}$ and a specific unoptimization operation set $\mS$, where $V$ is promised to be either an unoptimized quantum circuit $V=\mO(U, C)$ or $F(U\ket{0^n},V\ket{0^n}) \leq 1- 2/\poly(n)$. In other words, if the fidelity in Eq.~(\ref{eq: fidelity}) becomes 1, a recipe always exists.
The problem is defined to decide whether or not two quantum circuits, $U$ and $V$, are equivalent (unoptimized); 
If there exists a recipe $C$ such that $V=\mO(U, C)$ or equivalently if $F(U\ket{0^n},V\ket{0^n})=1$, 
the instance belongs to YES. Otherwise for any $C$, $F(U\ket{0^n},V\ket{0^n}) \leq 1- 2/\poly(n)$ and hence $V \neq \mO(U,C)$, the instance belongs to NO. 
\end{problem}

By the promise, this decision problem can be efficiently solved when using a quantum computer by measuring the fidelity $F(U\ket{0^n}, V\ket{0^n})$.
If this value is sufficiently close to 1 larger than the gap $1/\poly(n)$, we can decide the answer as YES. Otherwise, NO. 
Consequently, the promised quantum circuit equivalent test version is efficiently solvable on a quantum computer.
Furthermore, if a proof $C$ that unoptimizes $U$ into $V$ exists, we can efficiently verify that $V$ is an unoptimization of $U$ and decide the above problem. Therefore, the promised quantum circuit equivalence test also belongs to an NP problem.
Additionally, when the recipe generated from the unoptimization set $\mS$ is sufficiently complex, we can expect to create an equivalence test in Problem~\ref{Promised_equivalence_test} that is not trivially in P.
Note that if $V$ is randomly chosen as a random quantum circuit ~\cite{arute2019quantum}, $F(U\ket{0^n},V\ket{0^n})$ in Eq.~(\ref{eq: fidelity}) decays exponentially. Consequently, we can naturally construct $V$ that satisfies the promise.

The problems that belong to the intersection of NP and BQP, and whose belonging to P is nontrivial, are of great importance in demonstrating quantum advantage.
Within this class, there are a few problems proposed in prior research.
For instance, a prototypical problem is the prime factoring ~\cite{doi:10.1137/S0097539795293172}. 
Another nontrivial example is FH$_2$ (the second level of Fourier hierarchy) ~\cite{shi2005quantum,demarie2018classical},
where the BPP verifier can verify whether or not a given bit string has a sufficiently high probability.
Recently, it has been shown that there exist NP search problems that can be solved by BQP machines but not by BPP machines relative to a random oracle~\cite{yamakawa2022verifiable}.
The problem we have defined in this study is particularly valuable as it belongs to this limited complexity class.

Program obfuscation~\cite{10.1145/2160158.2160159} is also a concept similar to quantum circuit unoptimization.
Both primitives aim to make an input circuit complex while preserving its functionality.
Specifically, quantum obfuscation~\cite{alagic2016quantum} is required to conceal the quantum circuit's information, except for its input-output relationship, even from a quantum computer.
In contrast, quantum circuit unoptimization does not impose such strong requirements, as it only aims to create a redundant circuit that cannot be easily reversed to its original form.
While it is known that quantum obfuscation is difficult to construct in practice ~\cite{alagic2016quantum,obfuscation_difficulty_ananth,obfuscation_difficulty_alagic}, quantum circuit unoptimization is more feasible to implement.
Despite less stringent requirements, quantum circuit unoptimization has practical applications, such as generating benchmarks for compilers, as described in Sec.~\ref{sec: benchmark}.

\section{Compiler benchmark using quantum circuit unoptimization}
\label{sec: benchmark}

Unoptimized circuits can be applied not only to circuit equivalence tests but also to more practical aspects of quantum computing. The increase in circuit complexity caused by unoptimization can affect compilation efficiency and serves as a valuable metric for quantitatively comparing the performance of various quantum compilers.
In this section, as a specific application of unoptimization, we explain a method of compiler benchmark and compare the compilation performance between Qiskit and Pytket.

\subsection{Elementary operations for recipe}
\label{subsec: recipe}

We construct an unoptimization recipe $C$ using an elementary operation set $\mS$ consisting of the following four elements.

\begin{figure*}[htbp]
  \includegraphics[width=0.75\linewidth]{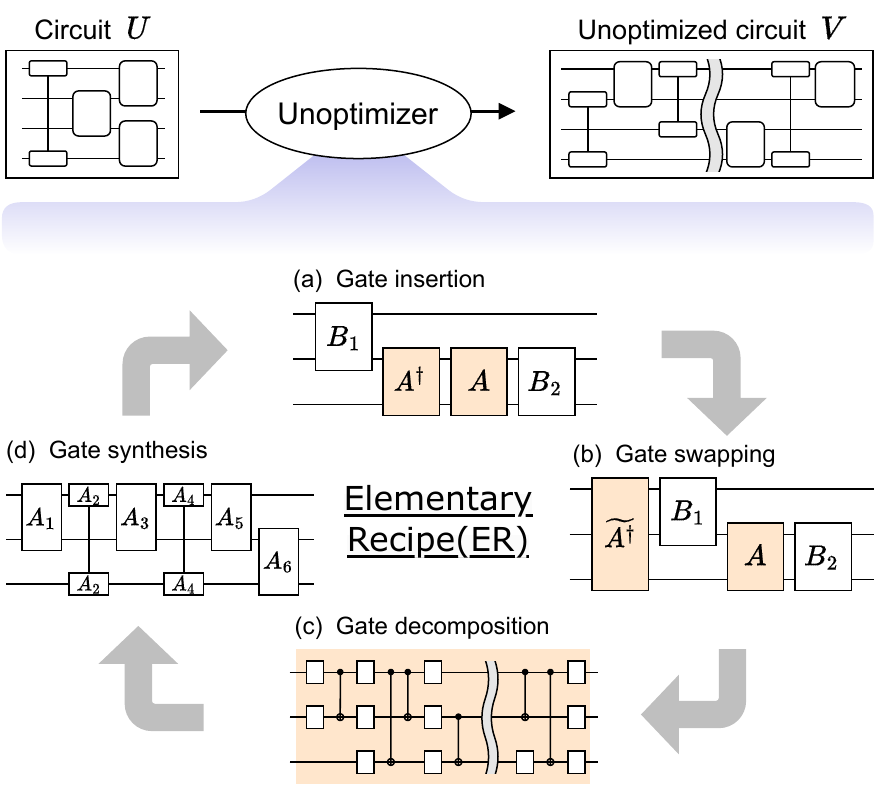}
  \caption{Quantum circuit unoptimization procedure that generates the unoptimized circuit $V$ from the original circuit $U$. The recipe $C$ we use in this study comprises four elements: {\it gate insertion, swapping, decomposition, and synthesis}. They make a circuit more redundant, keeping its equivalence. We call this cycle {\it an elementary recipe (ER)}.}
  \label{fig: recipe}
\end{figure*}

\begin{enumerate}
    \item {\it Gate insertion} (Fig.~\ref{fig: recipe}(a)): Places quantum gates in the quantum circuit.
    \item {\it Gate swapping} (Fig.~\ref{fig: recipe}(b)): Interchanges the positions of two quantum gates.
    \item {\it Gate decomposition} (Fig.~\ref{fig: recipe}(c)): Decomposes multi-qubit gates into the set of two-qubit or smaller gates
    using KAK decomposition~\cite{KAK, PhysRevA.69.010301, PhysRevA.69.062321}, Cosine-Sine Decomposition, and Quantum Shannon Decomposition~\cite{Shende_2006}.
    \item {\it Gate synthesis} (Fig.~\ref{fig: recipe}(d)): Greedy merges the quantum gates and generates two-qubit unitary blocks.
\end{enumerate}
Then we define the cycle of gate insertion, swapping, decomposition, and synthesis as {\it an elementary recipe (ER)}.

Suppose the original quantum circuit $U$ is given as a sequence of two-qubit gates, and their positions are randomly assigned~\cite{Moll_2018} as shown in Fig.~\ref{fig: recipe}.
ER starts with selecting a {\it pair} $\mP$ of two-qubit gates: $B_1$ and $B_2$ sharing only one common index. 
We insert an arbitrary two-qubit unitary gate $A$ and its Hermitian conjugate $A^{\dagger}$ between $B_1$ and $B_2$ as shown in Fig.~\ref{fig: recipe}(a). Since $AA^{\dagger}=I$, the equivalence of the circuit is preserved. 
However, they can be easily optimized by merging gates in the neighborhood. To prevent this, we apply {\it gate swapping} to $A^{\dagger}$ to obscure what the original circuit is. 
However, the circuit equivalence is not ensured by simply swapping $B_1$ and $A^{\dagger}$, 
so we need to replace $A^{\dagger}$ with a three-qubit unitary gate $\widetilde{A ^{\dagger}}$ as shown in Fig.~\ref{fig: recipe}(b) defined as
\begin{equation}
  \widetilde{A ^{\dagger}} =
  \left( B_1 \otimes I_1 \right) ^{\dagger}
  \left( I_1 \otimes A ^{\dagger} \right)
  \left( B_1 \otimes I_1 \right)
  \label{eq:calc_of_a_dagger_tilde}
\end{equation}
to cancel the byproduct that occurs after the swapping.
Finally $\widetilde{A ^{\dagger}}$ undergoes {\it decomposition} into the set of two-qubit or smaller gates (Fig.~\ref{fig: recipe}(c)) and {\it synthesize} the circuit such that the maximum block size of the gates is two (Fig.~\ref{fig: recipe}(d)).

In this way, ER can transform one pair of two-qubit gates into a more redundant sequence of unitary gates, keeping its circuit equivalence. Naturally, the more times we use ER, the longer the recipe $C$ becomes. In this study, we set an ER's iteration count $k$ as
\begin{equation}
  k= n^2.
  \label{eq:iteration_count}
\end{equation}
The reason why we choose $k$ in Eq.~(\ref{eq:iteration_count}) is that the total number of pairs in the $n$-qubit random circuit with a depth of $n$ is approximately $n^2$. This iteration count satisfies that $C$ is expressed as $\poly(n)$ classical bits as described in Sec.~\ref{sec: def_unopt}.

When we pick up a {\it pair} $\mP$ of two-qubit gates, we employ the following two pair-selection methods: $\mP_r, \mP_c$ and compare their differences. Naturally, the composition of recipe $C$ depends on them.
\renewcommand{\labelenumi}{(\Alph{enumi})}
\begin{itemize}
    \item {\it Random method} $\mP_r$: Randomly choose a two-qubit unitary gate. Search the candidate quantum gates that can implement ER as a pair with the target gate. When we get multiple choices, choose one randomly and then apply ER. If there is no candidate in the circuit, select the target gate again.
    \item {\it Concatenated method} $\mP_c$: First find a pair by $\mP_r$. After the second time, select the next pair consisting of a gate generated in the previous ER and an outside gate. When we get several options, choose the rightmost pair and then run ER. If there is no candidate in the circuit, randomly select a pair in the same way as $\mP_r$.
\end{itemize}

\subsection{Circuit generation and benchmark method}
\begin{figure}[t]
  \begin{center}
    \includegraphics[width=0.95\linewidth]{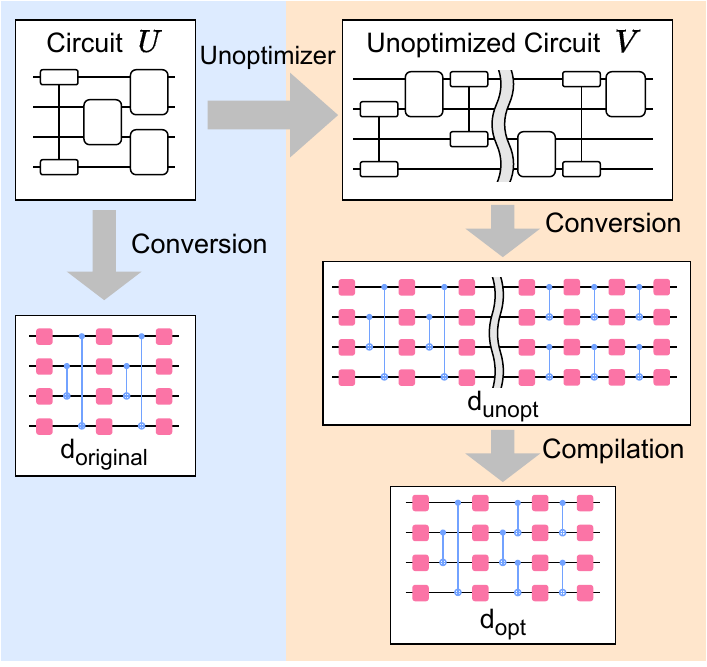}
    \caption{The overview of quantum compiler benchmark. The unoptimizer generates the unoptimized circuit $V$ from the original circuit $U$. These two circuits are converted into $\left\{ \mathrm{U3}, \mathrm{CX} \right\}$ gate sets (Qiskit objects), and we obtain the circuit depth $d_{\rm original}$ and $d_{\rm unopt}$. Compilers conduct the circuit reduction of $V$, and we get the optimized circuit whose depth is $d_{\rm opt}$. We measure it to evaluate compilation performance.}
    \label{fig: workflow_for_benchmark}
  \end{center}
\end{figure}

Unoptimized quantum circuits can be systematically generated by applying the recipe shown in Sec.~\ref{subsec: recipe} to the original circuit.
This study uses the constructed datasets to measure the circuit compression performance of two quantum compilers, Qiskit and Pytket. Fig.~\ref{fig: workflow_for_benchmark} displays the workflow of the compiler benchmark. Here, we explain the method of generating circuit datasets and assessing the performance of compilers in detail.

\renewcommand{\labelenumi}{(\roman{enumi})}
\vskip\baselineskip
\textbf{Generation of unoptimized circuit}
\begin{enumerate}
  \item Input a quantum circuit $U$ consisting of a sequence of two-qubit gates. The circuit depth is adjusted to the qubit count $n$ of $U$. 
  \item Construct a recipe $C$ using a method described in Sec.~\ref{subsec: recipe}. 
  \item Generate an unoptimized circuit $V$ from $U$ and $C$.
\end{enumerate}

\begin{figure}[t]
    \begin{algorithm}[H]
        \caption{Compiler benchmark}
        \begin{algorithmic}[1]
        \STATE \textbf{Generation of unoptimized circuit}
        \STATE \textbf{Input:} $U=$ $n$-qubit circuit, $\mP=\mP_r$ or $\mP_c$
        \STATE set $k=n^2$
        \STATE let $C[0, ... ,k-1]$ be a new array
        \FOR{$i \leftarrow 0$ to $k-1$}
        \STATE $C[i] \leftarrow$ $ \mP, ER$
        \ENDFOR
        \STATE $V \leftarrow \mO(U, C)$
        \STATE
        \STATE \textbf{Evaluation of compiler performance}
        \STATE $U, V \leftarrow U, V$ converted into $\left\{ \mathrm{U3}, \mathrm{CX} \right\}$ gate set
        \STATE $d_{\rm original}, d_{\rm unopt}=$ circuit depth of $U, V$
        \STATE $V \leftarrow V$ optimized by the compiler
        \STATE $d_{\rm opt}=$ circuit depth of $V$
        \end{algorithmic}
    \label{alg: benchmark}
    \end{algorithm}
\end{figure}

\vskip\baselineskip
\textbf{Evaluation of compiler performance}
\begin{itemize}
  \item Transform $U$ and $V$ into $\left\{ \mathrm{U3}, \mathrm{CX} \right\}$ Qiskit object gate set. $\mathrm{U3}$ means a single qubit rotation gate, and CX is $CNOT$ gate~\cite{Qiskit}. The obtained depth of the transformed circuits $U$ and $V$ are given as $d_{\rm original}$ and $d_{\rm unopt}$ respectively. Then we introduce the unoptimized ratio of them $r_{\rm unopt}$ defined as
    \begin{align}
          r_{\rm unopt} = \frac{d_{\rm unopt}}{d_{\rm original}},
          \label{eq:ratio_bef}
    \end{align}
    which is the value expected to grow according to $k$ in Eq.~(\ref{eq:iteration_count}). Our objective is to develop a quantum circuit unoptimization method that intentionally increases circuit complexity. In this context, the higher the value of $r_{\rm unopt}$, the superior the method is. 
  \item Evaluate the compression performance by compilers. The circuit is passed through the compiler, and its optimized depth $d_{\rm opt}$ is evaluated. Then we calculate the optimized ratio $r_{\rm opt}$ defined as
    \begin{align}
          r_{\rm opt} = \frac{d_{\rm opt}}{d_{\rm original}}.
          \label{eq:ratio_aft}
    \end{align}
\end{itemize}

\begin{figure*}[t]
    \begin{center}
      \includegraphics[width=1.0\linewidth]{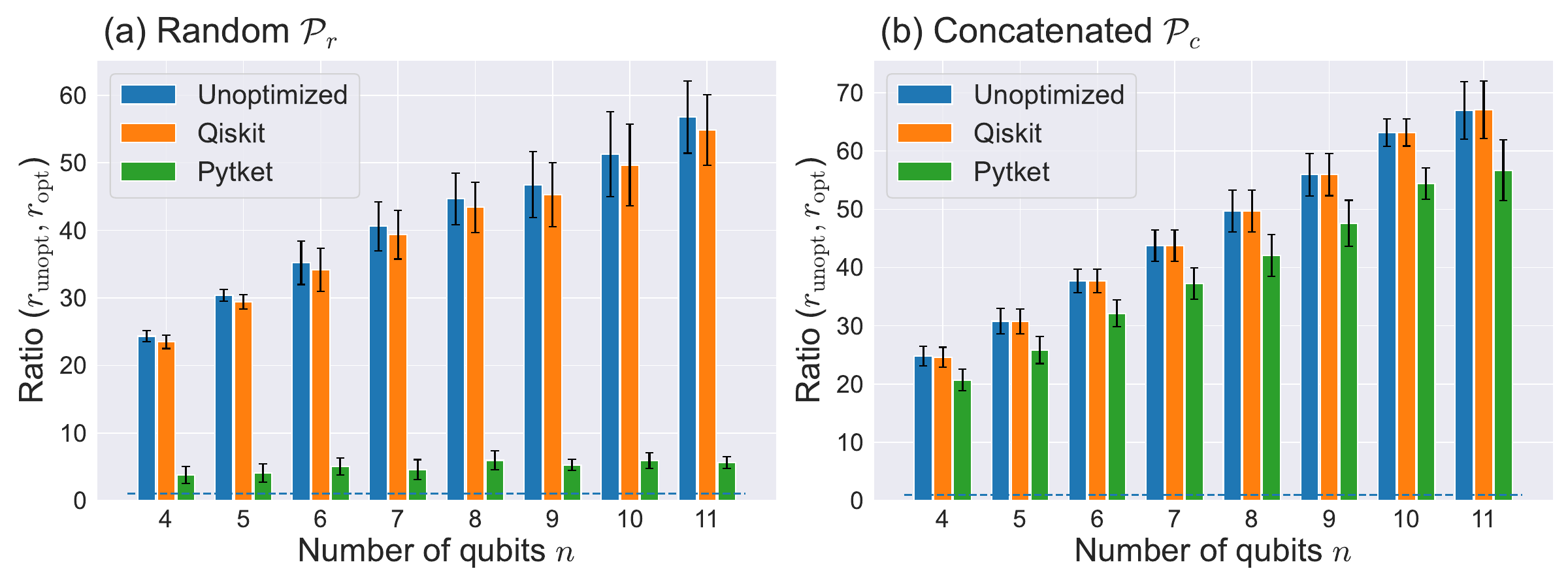}
    \end{center}
  \caption{Comparison with Qiskit and Pytket compiler performance for 30 random samples at each $n$. The vertical axis shows ratios defined in Eq.~(\ref{eq:ratio_bef}) and Eq.~(\ref{eq:ratio_aft}), and the horizontal one is the number of qubits. The dashed line represents a ratio of 1. If the input circuit has an optimal depth, the ratio will always be greater than 1. (a) The benchmark with a recipe generated by the random method $\mP_r$. (b) The benchmark with a recipe generated by the concatenated method $\mP_c$.}
  \label{fig: result_compiler}
\end{figure*}

\begin{figure*}[t]
  \begin{center}
    \includegraphics[width=0.8\linewidth]{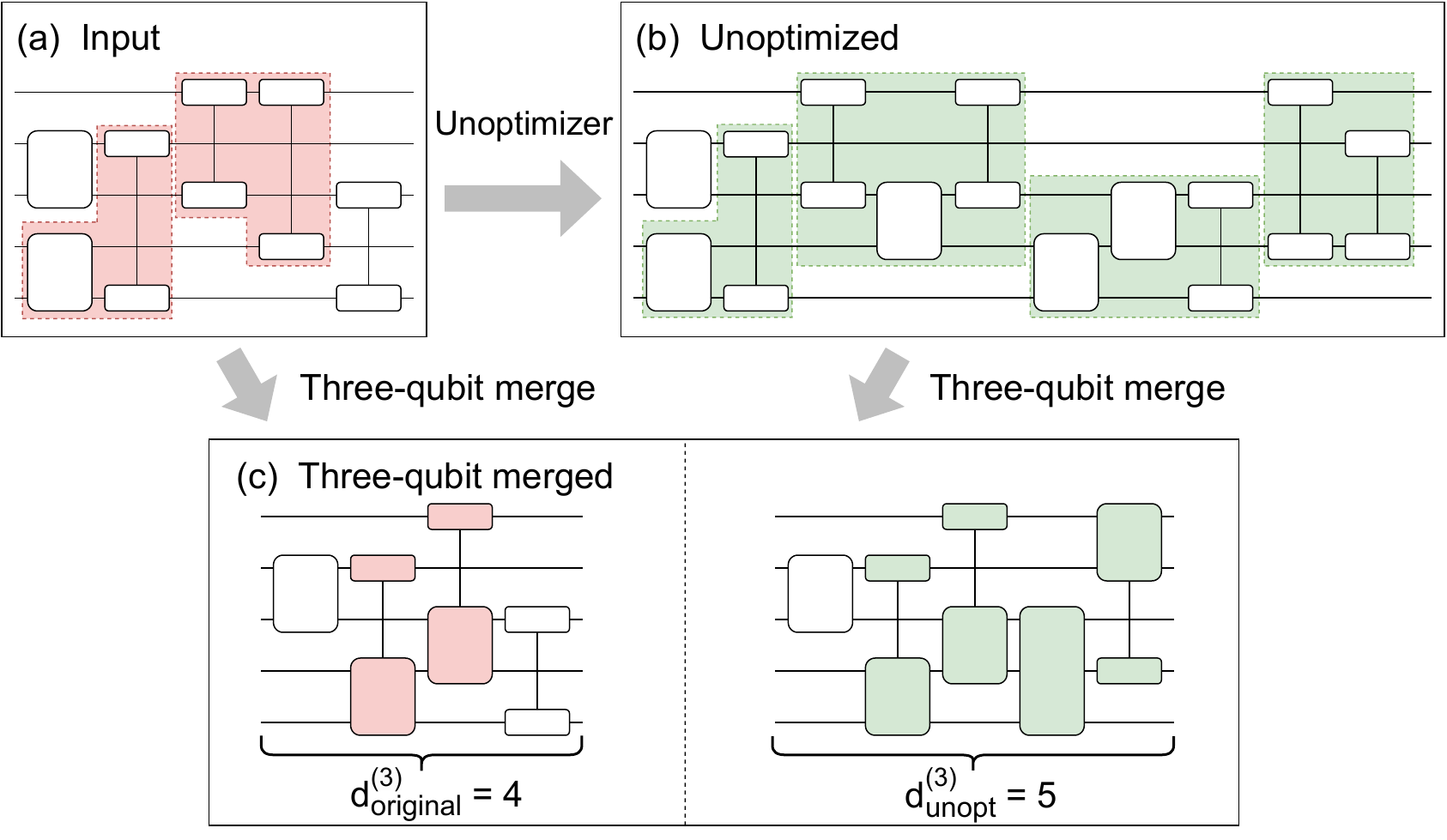}
    \caption{An example to convert the input and the unoptimized circuit into three-qubit merged ones. The obtained depth $d^{(3)}_{\rm original} $ and $d^{(3)}_{\rm unopt} $ are used for calculating $r^{(3)}_{\rm unopt} $ in Eq.~(\ref{eq:ratio_3qubit_merged}). This metric is designed to assess whether the unoptimization methods, $\mP_r$ (random method) and $\mP_c$ (concatenated method), bring local or non-local structural changes in the circuit. If $r^{(3)}_{\rm unopt}$ is close to~1, it suggests that unoptimization mainly affects fixed three-qubit regions, indicating a localized impact. In contrast, a larger $r^{(3)}_{\rm unopt}$ implies that the effect of unoptimization is widespread throughout the circuit.}
    \label{fig: 3qmerge_method}
  \end{center}
\end{figure*}

When we convert $U$ and $V$ into $\left\{ \mathrm{U3}, \mathrm{CX} \right\}$, we use \verb|Transpile(optimization_level=0)|, the module only performing circuit conversion without optimizations.
We compare the compression performance by two types of compilers: {\it Qiskit} \verb|Transpile(optimize_level=3)| and {\it Pytket} \verb|FullPeepholeOptimise|.
Compilers strive to minimize circuit depth as much as possible. Therefore, a more efficient compiler can bring the ratio $r_{\rm opt}$ closer to 1.
The decrease in value from $r_{\rm unopt}$ to $r_{\rm opt}$ corresponds to how well the compiler optimizes the complicated circuit. These ratios enable us to simultaneously assess the complexity level resulting from the unoptimization recipe and the performance of compilers.
Finally, an algorithm flow is shown in Algorithm~\ref{alg: benchmark}.

\subsection{Numerical experiments}

We examine the evolution of the ratios in Eq.~(\ref{eq:ratio_bef}) and Eq.~(\ref{eq:ratio_aft}) for quantum circuits with varying the number of qubits $n$ ranging from 4 to 11.
In this numerical simulation, we generate random two-qubit unitary quantum circuits using Qulacs~\cite{Qulacs}. Cirq~\cite{cirq_developers_2022_7465577} is utilized for two or three-qubit gate decomposition, and Qulacs conduct gate synthesis. The underlying code for this study is available on GitHub~\cite{qcunopt_mori}.

Fig.~\ref{fig: result_compiler} summarizes the results of our experiment. For each $n$, we generate 30 random circuits and show the average values and the standard deviations of $r_{\rm unopt}$ and $r_{\rm opt}$.
We verify the complexity level of quantum circuit unoptimization, $r_{\rm unopt}$, gradually grows with the increase in $n$, regardless of pair-selection methods.

In the case of the random method $\mP_r$, as shown in Fig.~\ref{fig: result_compiler}(a), 
$r_{\rm opt}$ of Qiskit is slightly smaller than the unoptimized ratio $r_{\rm unopt}$ in any number of qubits, meaning that Qiskit has slightly optimized the original circuit. In contrast, $r_{\rm opt}$ of Pytket considerably reduces the ratio to 5 or less regardless of the number of qubits. In other words, Pytket can significantly compress the circuit generated by the random method.
The concatenated method $\mP_c$ in Fig.~\ref{fig: result_compiler}(b), on the other hand, demonstrates that $r_{\rm opt}$ of both Qiskit and Pytket increases with the number of qubits. In other words, this method can generate the quantum circuit datasets that both compilers are challenging to optimize.

The difference between these two compilers is caused by the modules included in each compiler.
{\it Qiskit Transpile} employs \verb|UnitarySynthesis| to synthesize two-qubit gates and uses KAK decomposition for optimization.
In contrast, {\it Pytket FullPeeoholeOptimise} includes a module called \verb|ThreeQubitSquash|, which can synthesize the sequence of two-qubit gates positioned within three indices into a three-qubit gate and decompose it.
This module dramatically contributes to optimizing circuits generated by the random method. However, as shown in Fig.~\ref{fig: result_compiler}(b), the concatenated method generates a complicated circuit that Pytket cannot undo.

\begin{figure}[t]
  \begin{center}
    \includegraphics[width=1.0\linewidth]{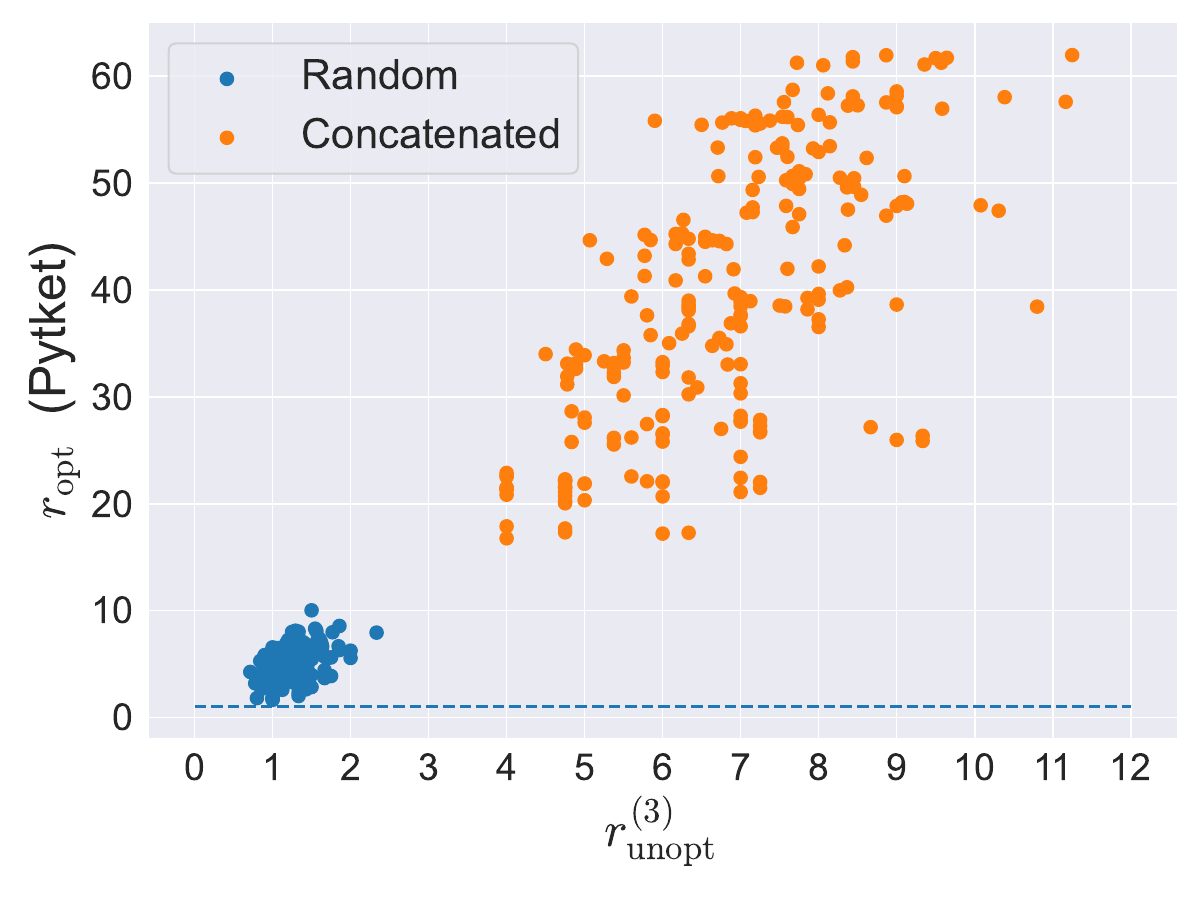}
    \caption{Comparison with two pair-selection methods: random and concatenated, using 30 random samples at each $n$ ranging from 4 to 11. The dashed line represents a ratio of 1. The higher value of $r^{(3)}_{\rm unopt} $ indicates that the generated sequences of unitary gates in the unoptimized circuit are effectively concatenated.}
    \label{fig: 3qmerge}
  \end{center}
\end{figure}

It is interesting to ask why the pair-selection method affects Pytket's compiler performance.
To address this question, we introduce a new operation to merge gates in the input circuit $U$ and the unoptimized circuit $V$ as shown in Fig.~\ref{fig: 3qmerge_method}.
We obtain a declined circuit depth when an unoptimized circuit merges into the three-qubit blocks. We define another ratio expressed as
\begin{align}
      r^{(3)}_{\rm unopt}  = 
      \frac{d^{(3)}_{\rm unopt} }{d^{(3)}_{\rm original} },
      \label{eq:ratio_3qubit_merged}
\end{align}
where $d^{(3)}_{\rm original}$ is the circuit depth obtained from the three-qubit merged circuit of $U$, and $d^{(3)}_{\rm unopt} $ is that of $V$ as shown in Fig.~\ref{fig: 3qmerge_method}(c).
One process of ER requires three qubits. If, when repeating ER, the sets of three-qubit indices used for ER remain unchanged, $r^{(3)}_{\rm unopt} $ in Eq.~(\ref{eq:ratio_3qubit_merged}) will be around 1. This is because the sequence of two-qubit gates positioned within three indices is merged into only one three-qubit block. Using this metric, we can examine the extent of variation in the three-qubit sets used for ER.

Fig.~\ref{fig: 3qmerge} displays the plot of $r^{(3)}_{\rm unopt} $ and $r_{\rm opt}$ of Pytket.
It can be seen that these two values have a positive correlation.
In the case of the random method, $r^{(3)}_{\rm unopt} $ stays around 1, indicating that the set of three-qubit indices used for ER remains the same.
On the other hand, the figure ranges from 4 to 11 in the concatenated method. This suggests that we need to change the qubit positions for ER so that they are effectively concatenated when we generate the complex circuits that Pytket's compiler is hard to optimize.

\section{Conclusion and Discussion}\label{sec: conclusion_and_discussion}

In this study, we propose quantum circuit unoptimization, a concept that transforms circuits into more redundant forms. Quantum circuit unoptimization offers intriguing possibilities in both theoretical and practical domains.
From a theoretical perspective, we use quantum circuit unoptimization to define decision problems as to whether two quantum circuits are equivalent or not.
One such problem is verifying circuit equivalence by checking the existence of a recipe that converts one circuit to the other, which is contained in the NP class.
Additionally, we combine this equivalence test with another that measures the inner product of the quantum states obtained from each circuit using a quantum computer, making the problem in both NP and BQP.
Given any unoptimization operation set, whether our proposed circuit equivalence test belongs to P or not remains an open question. If this problem is eventually proven to be in P, it signifies a remarkable simplicity in determining circuit equivalence despite the complexity of the circuit optimization problem~\cite{vandewetering2023optimising}.
Conversely, if one cannot find proof that this problem is in P, it is contained in the valuable complexity class, akin to prime factoring. Since our proposed problem can be solved by measuring the fidelity between quantum states, it is more feasible to implement this task in experiments than prime factoring.
We expect future work to uncover the detailed complexity of this test, like quantum random circuit sampling (RCS)~\cite{complexity_rcs,boixo2018characterizing,arute2019quantum}.

From the practical point of view, we focus on conducting benchmarks for evaluating compiler performance.
We construct a dataset of random quantum circuit sequences by combining four fundamental operations: gate insertion, swapping, decomposition, and synthesis, defined as Elementary Recipe (ER) in Fig.~\ref{fig: recipe}.
We quantitatively assess the performance of quantum circuit unoptimization and the compiler based on circuit depth ratios. Numerical experiments reveal that altering the recipe generation method of quantum circuit unoptimization results in variations in compiler compression effects.
Utilizing the simulations proposed in this study makes it possible to quantitatively evaluate the performance of unknown compilers and fairly compare them with others.
Furthermore, we expect this to contribute to future compiler improvements. For future works as benchmark tasks, we plan to conduct similar benchmarks for compilers such as Quartz~\cite{10.1145/3519939.3523433}, PyZX~\cite{Kissinger_2020},  VOQC~\cite{10.1145/3604630}, staq~\cite{Amy_2020}, and pyQuil~\cite{Smith_2020}.
Also, if the ER is constructed from a set that satisfies completeness~\cite{Cl_ment_2023, minimum_eq_theory}, in principle, any equivalent quantum circuit with a different description can be generated. Developing a recipe for dataset generation under such a setting would be an interesting direction for future research.

The versatility of quantum circuit unoptimization lies in its ability to systematically generate different quantum circuits while ensuring their equivalence as unitary operators. This property enables its application to a wide range of tasks.
Here, we propose some potential applications of quantum circuit unoptimization and aim to stimulate further research and development in the field of quantum computing.

Firstly, developing a protocol that utilizes an unoptimized circuit to verify whether the accessed server is quantum or classical would be intriguing.
Suppose $U$ is a quantum circuit that can be efficiently simulated classically, such as Clifford circuits, or circuits without large entangled blocks, meaning that we can efficiently compute its outcome with a sufficiently high probability. Then, we can generate its unoptimized circuit $V=\mO(U, C)$, send only $V$ to a server and request to sample the outputs. If finding the original circuit $U$ from $V$ is hard and the classical simulation of $V$ is intractable, it is difficult to provide correct outputs. In contrast, using a quantum computer, we can expect it to obtain the outcome even from $V$ efficiently. Therefore, we can test the quantum server using this property.
Alternatively, we can employ the promised version of the quantum circuit equivalence test described in Problem~\ref{Promised_equivalence_test}. The client sends classical descriptions of two quantum circuits, $U$ and $V$, generated as previously mentioned.
Then, the quantum server is asked whether or not $U$ and $V$ are equivalent.
The circuit rewriting technique derived from unoptimization can be employed to generate \textit{peaked circuits}, which have a high probability of producing a specific output on a computational basis. This is particularly relevant for verifiable quantum advantage experiments~\cite{aaronson_peaked}.

Secondly, we expect an unoptimizer to systematically generate a quantum advantageous dataset for quantum machine learning~\cite{Biamonte_2017}. Quantum machine learning is recognized as a field with significant potential to leverage the capabilities of quantum computers. Here, we focus on classifying many quantum circuit datasets based on the similarity of their output states.
Let us define a class of quantum circuits $\{A,B, ...\}$ represented by the unitary gates $\{U_A, U_B ,...\}$. The class refers to a set of quantum circuits that are unitarily equivalent to each other. Then, from each class, we can systematically generate $M$ samples of unoptimized circuits $\left\{ V^{(i)}_A, V^{(i)}_B,...\right\}_{i=1}^M$ via unoptimization recipes. The classical descriptions of these circuits are different; hence, it would be hard for a classical computer to classify this dataset into the classes. For the quantum computer, on the other hand, we can test them by measuring the fidelity of the outputs and then using the kernel-based classification method to infer the class~\cite{nakayama2023vqegenerated,placidi2023mnisq}.

Lastly, we explore the possibility of a quantum computer fidelity benchmark based on quantum circuit unoptimization.
RCS is a valuable approach to test the performance of a quantum computer~\cite{boixo2018characterizing,arute2019quantum}.
This task uses cross-entropy benchmarking computed from a random quantum circuit to indirectly estimate the circuit fidelity.
The unoptimization-based fidelity benchmark enables us to easily determine the ideal output from an original, simple quantum circuit. Therefore, we can directly estimate the circuit fidelity by comparing it with the results from an unoptimized circuit.
Suppose the original circuit $U$ is composed of an identity gate. Using an unoptimizer, we can generate an unoptimized circuit $V$ that is equivalent to $U$ by some elementary operations. Since $V$ consists of a series of elementary gates,
we can estimate how these gates work well on a real quantum computer by measuring the fidelity from the noisy unoptimized circuit $\mathcal{V}$, i.e., $\bra{0^n}\mathcal{V} \lbrack \ket{0^n}\bra{0^n} \rbrack \ket{0^n}$, and investigating how much the value deviates from 1.
In this case, $V$ would be run as is without optimization. Additionally, the unoptimization-based method, with its unique ability to control the circuit structure and complexity through flexible changes in the unoptimization recipe, offers a distinct advantage over other fidelity benchmarks. It can construct completely disordered circuits that still result in an overall identity. It allows us to see that the unoptimization-based method is capable of capturing a broader range of noise, thereby providing a more comprehensive evaluation of the quantum computer's performance.

\vskip\baselineskip

\section*{Acknowledgement}
The authors acknowledge helpful discussions and support by F. Le Gall and M. Miyamoto. This work is supported by the MEXT Quantum Leap Flagship Program (MEXT Q-LEAP) Grants No. JPMXS0118067394 and No. JPMXS0120319794, JST COI-NEXT Grant No. JPMJPF2014, and JSPS KAKENHI Grant No. JP25KJ1712.

\bibliography{bib.bib}

\begin{thebibliography}{44}%
\makeatletter
\providecommand \@ifxundefined [1]{%
 \@ifx{#1\undefined}
}%
\providecommand \@ifnum [1]{%
 \ifnum #1\expandafter \@firstoftwo
 \else \expandafter \@secondoftwo
 \fi
}%
\providecommand \@ifx [1]{%
 \ifx #1\expandafter \@firstoftwo
 \else \expandafter \@secondoftwo
 \fi
}%
\providecommand \natexlab [1]{#1}%
\providecommand \enquote  [1]{``#1''}%
\providecommand \bibnamefont  [1]{#1}%
\providecommand \bibfnamefont [1]{#1}%
\providecommand \citenamefont [1]{#1}%
\providecommand \href@noop [0]{\@secondoftwo}%
\providecommand \href [0]{\begingroup \@sanitize@url \@href}%
\providecommand \@href[1]{\@@startlink{#1}\@@href}%
\providecommand \@@href[1]{\endgroup#1\@@endlink}%
\providecommand \@sanitize@url [0]{\catcode `\\12\catcode `\$12\catcode `\&12\catcode `\#12\catcode `\^12\catcode `\_12\catcode `\%12\relax}%
\providecommand \@@startlink[1]{}%
\providecommand \@@endlink[0]{}%
\providecommand \url  [0]{\begingroup\@sanitize@url \@url }%
\providecommand \@url [1]{\endgroup\@href {#1}{\urlprefix }}%
\providecommand \urlprefix  [0]{URL }%
\providecommand \Eprint [0]{\href }%
\providecommand \doibase [0]{https://doi.org/}%
\providecommand \selectlanguage [0]{\@gobble}%
\providecommand \bibinfo  [0]{\@secondoftwo}%
\providecommand \bibfield  [0]{\@secondoftwo}%
\providecommand \translation [1]{[#1]}%
\providecommand \BibitemOpen [0]{}%
\providecommand \bibitemStop [0]{}%
\providecommand \bibitemNoStop [0]{.\EOS\space}%
\providecommand \EOS [0]{\spacefactor3000\relax}%
\providecommand \BibitemShut  [1]{\csname bibitem#1\endcsname}%
\let\auto@bib@innerbib\@empty
\bibitem [{\citenamefont {Micheli}(1994)}]{micheli1994synthesis}%
  \BibitemOpen
  \bibfield  {author} {\bibinfo {author} {\bibfnamefont {G.~D.}\ \bibnamefont {Micheli}},\ }\href@noop {} {\emph {\bibinfo {title} {Synthesis and Optimization of Digital Circuits}}},\ \bibinfo {edition} {1st}\ ed.\ (\bibinfo  {publisher} {McGraw-Hill Higher Education, Blacklick, OH},\ \bibinfo {year} {1994})\BibitemShut {NoStop}%
\bibitem [{\citenamefont {Cook}(1971)}]{cook1971complexity}%
  \BibitemOpen
  \bibfield  {author} {\bibinfo {author} {\bibfnamefont {S.~A.}\ \bibnamefont {Cook}},\ }\bibfield  {title} {\bibinfo {title} {The complexity of theorem-proving procedures},\ }in\ \href {https://doi.org/10.1145/800157.805047} {\emph {\bibinfo {booktitle} {Proceedings of the Third Annual ACM Symposium on Theory of Computing}}},\ \bibinfo {series and number} {STOC '71}\ (\bibinfo  {publisher} {Association for Computing Machinery},\ \bibinfo {address} {New York},\ \bibinfo {year} {1971})\ pp.\ \bibinfo {pages} {151--158}\BibitemShut {NoStop}%
\bibitem [{\citenamefont {Levin}(1973)}]{levin1973universal}%
  \BibitemOpen
  \bibfield  {author} {\bibinfo {author} {\bibfnamefont {L.~A.}\ \bibnamefont {Levin}},\ }\bibfield  {title} {\bibinfo {title} {Universal sequential search problems},\ }\href {http://mi.mathnet.ru/ppi914} {\bibfield  {journal} {\bibinfo  {journal} {Probl. Peredachi Inf.}\ }\textbf {\bibinfo {volume} {9}},\ \bibinfo {pages} {115} (\bibinfo {year} {1973})}\BibitemShut {NoStop}%
\bibitem [{\citenamefont {Trakhtenbrot}(1984)}]{trakhtenbrot1984survey}%
  \BibitemOpen
  \bibfield  {author} {\bibinfo {author} {\bibfnamefont {B.}~\bibnamefont {Trakhtenbrot}},\ }\bibfield  {title} {\bibinfo {title} {A survey of russian approaches to perebor (brute-force searches) algorithms},\ }\href {https://doi.org/10.1109/MAHC.1984.10036} {\bibfield  {journal} {\bibinfo  {journal} {Annals of the History of Computing}\ }\textbf {\bibinfo {volume} {6}},\ \bibinfo {pages} {384} (\bibinfo {year} {1984})}\BibitemShut {NoStop}%
\bibitem [{\citenamefont {Ilango}\ \emph {et~al.}(2020)\citenamefont {Ilango}, \citenamefont {Loff},\ and\ \citenamefont {Oliveira}}]{ilango2020np}%
  \BibitemOpen
  \bibfield  {author} {\bibinfo {author} {\bibfnamefont {R.}~\bibnamefont {Ilango}}, \bibinfo {author} {\bibfnamefont {B.}~\bibnamefont {Loff}},\ and\ \bibinfo {author} {\bibfnamefont {I.~C.}\ \bibnamefont {Oliveira}},\ }\bibfield  {title} {\bibinfo {title} {{NP}-hardness of circuit minimization for multi-output functions},\ }in\ \href {https://doi.org/10.4230/LIPIcs.CCC.2020.22} {\emph {\bibinfo {booktitle} {Proceedings of the 35th Computational Complexity Conference}}},\ \bibinfo {series and number} {CCC '20}\ (\bibinfo  {publisher} {Schloss Dagstuhl--Leibniz-Zentrum f{\"u}r Informatik},\ \bibinfo {address} {Dagstuhl, DEU},\ \bibinfo {year} {2020})\BibitemShut {NoStop}%
\bibitem [{\citenamefont {Preskill}(2018)}]{preskill2018quantum}%
  \BibitemOpen
  \bibfield  {author} {\bibinfo {author} {\bibfnamefont {J.}~\bibnamefont {Preskill}},\ }\bibfield  {title} {\bibinfo {title} {Quantum computing in the nisq era and beyond},\ }\href {https://doi.org/10.22331/q-2018-08-06-79} {\bibfield  {journal} {\bibinfo  {journal} {Quantum}\ }\textbf {\bibinfo {volume} {2}},\ \bibinfo {pages} {79} (\bibinfo {year} {2018})}\BibitemShut {NoStop}%
\bibitem [{\citenamefont {Brandhofer}\ \emph {et~al.}(2021)\citenamefont {Brandhofer}, \citenamefont {Devitt}, \citenamefont {Wellens},\ and\ \citenamefont {Polian}}]{brandhofer2021special}%
  \BibitemOpen
  \bibfield  {author} {\bibinfo {author} {\bibfnamefont {S.}~\bibnamefont {Brandhofer}}, \bibinfo {author} {\bibfnamefont {S.}~\bibnamefont {Devitt}}, \bibinfo {author} {\bibfnamefont {T.}~\bibnamefont {Wellens}},\ and\ \bibinfo {author} {\bibfnamefont {I.}~\bibnamefont {Polian}},\ }\bibfield  {title} {\bibinfo {title} {Special session: Noisy intermediate-scale quantum (nisq) computers―how they work, how they fail, how to test them?},\ }in\ \href {https://doi.org/10.1109/VTS50974.2021.9441047} {\emph {\bibinfo {booktitle} {2021 IEEE 39th VLSI Test Symposium (VTS)}}}\ (\bibinfo {address} {IEEE, New York},\ \bibinfo {year} {2021})\ pp.\ \bibinfo {pages} {1--10}\BibitemShut {NoStop}%
\bibitem [{\citenamefont {Xu}\ \emph {et~al.}(2022)\citenamefont {Xu}, \citenamefont {Li}, \citenamefont {Padon}, \citenamefont {Lin}, \citenamefont {Pointing}, \citenamefont {Hirth}, \citenamefont {Ma}, \citenamefont {Palsberg}, \citenamefont {Aiken}, \citenamefont {Acar},\ and\ \citenamefont {Jia}}]{10.1145/3519939.3523433}%
  \BibitemOpen
  \bibfield  {author} {\bibinfo {author} {\bibfnamefont {M.}~\bibnamefont {Xu}}, \bibinfo {author} {\bibfnamefont {Z.}~\bibnamefont {Li}}, \bibinfo {author} {\bibfnamefont {O.}~\bibnamefont {Padon}}, \bibinfo {author} {\bibfnamefont {S.}~\bibnamefont {Lin}}, \bibinfo {author} {\bibfnamefont {J.}~\bibnamefont {Pointing}}, \bibinfo {author} {\bibfnamefont {A.}~\bibnamefont {Hirth}}, \bibinfo {author} {\bibfnamefont {H.}~\bibnamefont {Ma}}, \bibinfo {author} {\bibfnamefont {J.}~\bibnamefont {Palsberg}}, \bibinfo {author} {\bibfnamefont {A.}~\bibnamefont {Aiken}}, \bibinfo {author} {\bibfnamefont {U.~A.}\ \bibnamefont {Acar}},\ and\ \bibinfo {author} {\bibfnamefont {Z.}~\bibnamefont {Jia}},\ }\bibfield  {title} {\bibinfo {title} {Quartz: Superoptimization of quantum circuits},\ }in\ \href {https://doi.org/10.1145/3519939.3523433} {\emph {\bibinfo {booktitle} {Proceedings of the 43rd ACM SIGPLAN International Conference on Programming Language Design and Implementation}}},\ \bibinfo {series and number} {PLDI
  2022}\ (\bibinfo  {publisher} {Association for Computing Machinery},\ \bibinfo {address} {New York},\ \bibinfo {year} {2022})\ pp.\ \bibinfo {pages} {625--640}\BibitemShut {NoStop}%
\bibitem [{\citenamefont {Kharkov}\ \emph {et~al.}(2022)\citenamefont {Kharkov}, \citenamefont {Ivanova}, \citenamefont {Mikhantiev},\ and\ \citenamefont {Kotelnikov}}]{kharkov2022arline}%
  \BibitemOpen
  \bibfield  {author} {\bibinfo {author} {\bibfnamefont {Y.}~\bibnamefont {Kharkov}}, \bibinfo {author} {\bibfnamefont {A.}~\bibnamefont {Ivanova}}, \bibinfo {author} {\bibfnamefont {E.}~\bibnamefont {Mikhantiev}},\ and\ \bibinfo {author} {\bibfnamefont {A.}~\bibnamefont {Kotelnikov}},\ }\href@noop {} {\bibinfo {title} {Arline benchmarks: Automated benchmarking platform for quantum compilers}} (\bibinfo {year} {2022}),\ \Eprint {https://arxiv.org/abs/2202.14025} {arXiv:2202.14025 [quant-ph]} \BibitemShut {NoStop}%
\bibitem [{\citenamefont {Cl\'{e}ment}\ \emph {et~al.}(2023)\citenamefont {Cl\'{e}ment}, \citenamefont {Heurtel}, \citenamefont {Mansfield}, \citenamefont {Perdrix},\ and\ \citenamefont {Valiron}}]{Cl_ment_2023}%
  \BibitemOpen
  \bibfield  {author} {\bibinfo {author} {\bibfnamefont {A.}~\bibnamefont {Cl\'{e}ment}}, \bibinfo {author} {\bibfnamefont {N.}~\bibnamefont {Heurtel}}, \bibinfo {author} {\bibfnamefont {S.}~\bibnamefont {Mansfield}}, \bibinfo {author} {\bibfnamefont {S.}~\bibnamefont {Perdrix}},\ and\ \bibinfo {author} {\bibfnamefont {B.}~\bibnamefont {Valiron}},\ }\bibfield  {title} {\bibinfo {title} {A complete equational theory for quantum circuits},\ }in\ \href {https://doi.org/10.1109/lics56636.2023.10175801} {\emph {\bibinfo {booktitle} {2023 38th Annual ACM/IEEE Symposium on Logic in Computer Science (LICS)}}}\ (\bibinfo  {publisher} {IEEE},\ \bibinfo {year} {2023})\ pp.\ \bibinfo {pages} {1--13}\BibitemShut {NoStop}%
\bibitem [{\citenamefont {Cl\'{e}ment}\ \emph {et~al.}(2024)\citenamefont {Cl\'{e}ment}, \citenamefont {Delorme},\ and\ \citenamefont {Perdrix}}]{minimum_eq_theory}%
  \BibitemOpen
  \bibfield  {author} {\bibinfo {author} {\bibfnamefont {A.}~\bibnamefont {Cl\'{e}ment}}, \bibinfo {author} {\bibfnamefont {N.}~\bibnamefont {Delorme}},\ and\ \bibinfo {author} {\bibfnamefont {S.}~\bibnamefont {Perdrix}},\ }\bibfield  {title} {\bibinfo {title} {Minimal equational theories for quantum circuits},\ }in\ \href {https://doi.org/10.1145/3661814.3662088} {\emph {\bibinfo {booktitle} {Proceedings of the 39th Annual ACM/IEEE Symposium on Logic in Computer Science}}},\ \bibinfo {series and number} {LICS '24}\ (\bibinfo  {publisher} {Association for Computing Machinery},\ \bibinfo {address} {New York},\ \bibinfo {year} {2024})\BibitemShut {NoStop}%
\bibitem [{\citenamefont {Treinish}\ \emph {et~al.}(2022)\citenamefont {Treinish}, \citenamefont {Gambetta}, \citenamefont {Rodr\'{i}guez}, \citenamefont {Bello}, \citenamefont {Marques}, \citenamefont {Wood}, \citenamefont {Gacon}, \citenamefont {Nation}, \citenamefont {Gomez}, \citenamefont {ewinston}, \citenamefont {Cross}, \citenamefont {Lishman}, \citenamefont {Krsulich}, \citenamefont {Sertage}, \citenamefont {Wood}, \citenamefont {Alexander}, \citenamefont {Capelluto}, \citenamefont {Kanazawa}, \citenamefont {de~la Puente~Gonz\'{a}lez}, \citenamefont {Hamamura}, \citenamefont {Navarro}, \citenamefont {Rubio}, \citenamefont {Imamichi}, \citenamefont {Tod}, \citenamefont {Mezzacapo}, \citenamefont {Bishop}, \citenamefont {Hu}, \citenamefont {abbycross}, \citenamefont {Zoufalc},\ and\ \citenamefont {Chen}}]{Qiskit}%
  \BibitemOpen
  \bibfield  {author} {\bibinfo {author} {\bibfnamefont {M.}~\bibnamefont {Treinish}}, \bibinfo {author} {\bibfnamefont {J.}~\bibnamefont {Gambetta}}, \bibinfo {author} {\bibfnamefont {D.~M.}\ \bibnamefont {Rodr\'{i}guez}}, \bibinfo {author} {\bibfnamefont {L.}~\bibnamefont {Bello}}, \bibinfo {author} {\bibfnamefont {M.}~\bibnamefont {Marques}}, \bibinfo {author} {\bibfnamefont {C.~J.}\ \bibnamefont {Wood}}, \bibinfo {author} {\bibfnamefont {J.}~\bibnamefont {Gacon}}, \bibinfo {author} {\bibfnamefont {P.}~\bibnamefont {Nation}}, \bibinfo {author} {\bibfnamefont {J.}~\bibnamefont {Gomez}}, \bibinfo {author} {\bibnamefont {ewinston}}, \bibinfo {author} {\bibfnamefont {A.}~\bibnamefont {Cross}}, \bibinfo {author} {\bibfnamefont {J.}~\bibnamefont {Lishman}}, \bibinfo {author} {\bibfnamefont {K.}~\bibnamefont {Krsulich}}, \bibinfo {author} {\bibfnamefont {I.~F.}\ \bibnamefont {Sertage}}, \bibinfo {author} {\bibfnamefont {S.}~\bibnamefont {Wood}}, \bibinfo {author} {\bibfnamefont {T.}~\bibnamefont {Alexander}},
  \bibinfo {author} {\bibfnamefont {L.}~\bibnamefont {Capelluto}}, \bibinfo {author} {\bibfnamefont {N.}~\bibnamefont {Kanazawa}}, \bibinfo {author} {\bibfnamefont {S.}~\bibnamefont {de~la Puente~Gonz\'{a}lez}}, \bibinfo {author} {\bibfnamefont {I.}~\bibnamefont {Hamamura}}, \bibinfo {author} {\bibfnamefont {E.}~\bibnamefont {Navarro}}, \bibinfo {author} {\bibfnamefont {J.}~\bibnamefont {Rubio}}, \bibinfo {author} {\bibfnamefont {T.}~\bibnamefont {Imamichi}}, \bibinfo {author} {\bibfnamefont {M.}~\bibnamefont {Tod}}, \bibinfo {author} {\bibfnamefont {A.}~\bibnamefont {Mezzacapo}}, \bibinfo {author} {\bibfnamefont {L.}~\bibnamefont {Bishop}}, \bibinfo {author} {\bibfnamefont {S.}~\bibnamefont {Hu}}, \bibinfo {author} {\bibnamefont {abbycross}}, \bibinfo {author} {\bibnamefont {Zoufalc}},\ and\ \bibinfo {author} {\bibfnamefont {R.}~\bibnamefont {Chen}},\ }\href {https://doi.org/10.5281/zenodo.7352457} {\bibinfo {title} {Qiskit/qiskit-terra: Qiskit terra 0.22.3}} (\bibinfo {year} {2022})\BibitemShut {NoStop}%
\bibitem [{\citenamefont {Sivarajah}\ \emph {et~al.}(2021)\citenamefont {Sivarajah}, \citenamefont {Dilkes}, \citenamefont {Cowtan}, \citenamefont {Simmons}, \citenamefont {Edgington},\ and\ \citenamefont {Duncan}}]{Pytket}%
  \BibitemOpen
  \bibfield  {author} {\bibinfo {author} {\bibfnamefont {S.}~\bibnamefont {Sivarajah}}, \bibinfo {author} {\bibfnamefont {S.}~\bibnamefont {Dilkes}}, \bibinfo {author} {\bibfnamefont {A.}~\bibnamefont {Cowtan}}, \bibinfo {author} {\bibfnamefont {W.}~\bibnamefont {Simmons}}, \bibinfo {author} {\bibfnamefont {A.}~\bibnamefont {Edgington}},\ and\ \bibinfo {author} {\bibfnamefont {R.}~\bibnamefont {Duncan}},\ }\bibfield  {title} {\bibinfo {title} {t|ket$\rangle$: a retargetable compiler for nisq devices},\ }\href {https://doi.org/10.1088/2058-9565/ab8e92} {\bibfield  {journal} {\bibinfo  {journal} {Quantum Science and Technology}\ }\textbf {\bibinfo {volume} {6}},\ \bibinfo {pages} {014003} (\bibinfo {year} {2021})}\BibitemShut {NoStop}%
\bibitem [{\citenamefont {Janzing}\ \emph {et~al.}(2005)\citenamefont {Janzing}, \citenamefont {Wocjan},\ and\ \citenamefont {Beth}}]{qma_complete}%
  \BibitemOpen
  \bibfield  {author} {\bibinfo {author} {\bibfnamefont {D.}~\bibnamefont {Janzing}}, \bibinfo {author} {\bibfnamefont {P.}~\bibnamefont {Wocjan}},\ and\ \bibinfo {author} {\bibfnamefont {T.}~\bibnamefont {Beth}},\ }\bibfield  {title} {\bibinfo {title} {Non-identity-check is {QMA}-complete,},\ }\href {https://doi.org/10.1142/S0219749905001067} {\bibfield  {journal} {\bibinfo  {journal} {International Journal of Quantum Information}\ }\textbf {\bibinfo {volume} {03}},\ \bibinfo {pages} {463} (\bibinfo {year} {2005})},\ \Eprint {https://arxiv.org/abs/https://doi.org/10.1142/S0219749905001067} {https://doi.org/10.1142/S0219749905001067} \BibitemShut {NoStop}%
\bibitem [{\citenamefont {Rethinasamy}\ \emph {et~al.}(2023)\citenamefont {Rethinasamy}, \citenamefont {Agarwal}, \citenamefont {Sharma},\ and\ \citenamefont {Wilde}}]{pure_pure_fidelity}%
  \BibitemOpen
  \bibfield  {author} {\bibinfo {author} {\bibfnamefont {S.}~\bibnamefont {Rethinasamy}}, \bibinfo {author} {\bibfnamefont {R.}~\bibnamefont {Agarwal}}, \bibinfo {author} {\bibfnamefont {K.}~\bibnamefont {Sharma}},\ and\ \bibinfo {author} {\bibfnamefont {M.~M.}\ \bibnamefont {Wilde}},\ }\bibfield  {title} {\bibinfo {title} {Estimating distinguishability measures on quantum computers},\ }\href {https://doi.org/10.1103/PhysRevA.108.012409} {\bibfield  {journal} {\bibinfo  {journal} {Phys. Rev. A}\ }\textbf {\bibinfo {volume} {108}},\ \bibinfo {pages} {012409} (\bibinfo {year} {2023})}\BibitemShut {NoStop}%
\bibitem [{\citenamefont {Cai}\ \emph {et~al.}(2023)\citenamefont {Cai}, \citenamefont {Babbush}, \citenamefont {Benjamin}, \citenamefont {Endo}, \citenamefont {Huggins}, \citenamefont {Li}, \citenamefont {McClean},\ and\ \citenamefont {O'Brien}}]{mitigation}%
  \BibitemOpen
  \bibfield  {author} {\bibinfo {author} {\bibfnamefont {Z.}~\bibnamefont {Cai}}, \bibinfo {author} {\bibfnamefont {R.}~\bibnamefont {Babbush}}, \bibinfo {author} {\bibfnamefont {S.~C.}\ \bibnamefont {Benjamin}}, \bibinfo {author} {\bibfnamefont {S.}~\bibnamefont {Endo}}, \bibinfo {author} {\bibfnamefont {W.~J.}\ \bibnamefont {Huggins}}, \bibinfo {author} {\bibfnamefont {Y.}~\bibnamefont {Li}}, \bibinfo {author} {\bibfnamefont {J.~R.}\ \bibnamefont {McClean}},\ and\ \bibinfo {author} {\bibfnamefont {T.~E.}\ \bibnamefont {O'Brien}},\ }\bibfield  {title} {\bibinfo {title} {Quantum error mitigation},\ }\href {https://doi.org/10.1103/RevModPhys.95.045005} {\bibfield  {journal} {\bibinfo  {journal} {Rev. Mod. Phys.}\ }\textbf {\bibinfo {volume} {95}},\ \bibinfo {pages} {045005} (\bibinfo {year} {2023})}\BibitemShut {NoStop}%
\bibitem [{\citenamefont {Arute}\ \emph {et~al.}(2019)\citenamefont {Arute}, \citenamefont {Arya}, \citenamefont {Babbush}, \citenamefont {Bacon}, \citenamefont {Bardin}, \citenamefont {Barends}, \citenamefont {Biswas}, \citenamefont {Boixo}, \citenamefont {Brandao}, \citenamefont {Buell}, \citenamefont {Burkett}, \citenamefont {Chen}, \citenamefont {Chen}, \citenamefont {Chiaro}, \citenamefont {Collins}, \citenamefont {Courtney}, \citenamefont {Dunsworth}, \citenamefont {Farhi}, \citenamefont {Foxen}, \citenamefont {Fowler}, \citenamefont {Gidney}, \citenamefont {Giustina}, \citenamefont {Graff}, \citenamefont {Guerin}, \citenamefont {Habegger}, \citenamefont {Harrigan}, \citenamefont {Hartmann}, \citenamefont {Ho}, \citenamefont {Hoffmann}, \citenamefont {Huang}, \citenamefont {Humble}, \citenamefont {Isakov}, \citenamefont {Jeffrey}, \citenamefont {Jiang}, \citenamefont {Kafri}, \citenamefont {Kechedzhi}, \citenamefont {Kelly}, \citenamefont {Klimov}, \citenamefont {Knysh}, \citenamefont {Korotkov},
  \citenamefont {Kostritsa}, \citenamefont {Landhuis}, \citenamefont {Lindmark}, \citenamefont {Lucero}, \citenamefont {Lyakh}, \citenamefont {Mandr\`{a}}, \citenamefont {McClean}, \citenamefont {McEwen}, \citenamefont {Megrant}, \citenamefont {Mi}, \citenamefont {Michielsen}, \citenamefont {Mohseni}, \citenamefont {Mutus}, \citenamefont {Naaman}, \citenamefont {Neeley}, \citenamefont {Neill}, \citenamefont {Niu}, \citenamefont {Ostby}, \citenamefont {Petukhov}, \citenamefont {Platt}, \citenamefont {Quintana}, \citenamefont {Rieffel}, \citenamefont {Roushan}, \citenamefont {Rubin}, \citenamefont {Sank}, \citenamefont {Satzinger}, \citenamefont {Smelyanskiy}, \citenamefont {Sung}, \citenamefont {Trevithick}, \citenamefont {Vainsencher}, \citenamefont {Villalonga}, \citenamefont {White}, \citenamefont {Yao}, \citenamefont {Yeh}, \citenamefont {Zalcman}, \citenamefont {Neven},\ and\ \citenamefont {Martinis}}]{arute2019quantum}%
  \BibitemOpen
  \bibfield  {author} {\bibinfo {author} {\bibfnamefont {F.}~\bibnamefont {Arute}}, \bibinfo {author} {\bibfnamefont {K.}~\bibnamefont {Arya}}, \bibinfo {author} {\bibfnamefont {R.}~\bibnamefont {Babbush}}, \bibinfo {author} {\bibfnamefont {D.}~\bibnamefont {Bacon}}, \bibinfo {author} {\bibfnamefont {J.~C.}\ \bibnamefont {Bardin}}, \bibinfo {author} {\bibfnamefont {R.}~\bibnamefont {Barends}}, \bibinfo {author} {\bibfnamefont {R.}~\bibnamefont {Biswas}}, \bibinfo {author} {\bibfnamefont {S.}~\bibnamefont {Boixo}}, \bibinfo {author} {\bibfnamefont {F.~G. S.~L.}\ \bibnamefont {Brandao}}, \bibinfo {author} {\bibfnamefont {D.~A.}\ \bibnamefont {Buell}}, \bibinfo {author} {\bibfnamefont {B.}~\bibnamefont {Burkett}}, \bibinfo {author} {\bibfnamefont {Y.}~\bibnamefont {Chen}}, \bibinfo {author} {\bibfnamefont {Z.}~\bibnamefont {Chen}}, \bibinfo {author} {\bibfnamefont {B.}~\bibnamefont {Chiaro}}, \bibinfo {author} {\bibfnamefont {R.}~\bibnamefont {Collins}}, \bibinfo {author} {\bibfnamefont {W.}~\bibnamefont
  {Courtney}}, \bibinfo {author} {\bibfnamefont {A.}~\bibnamefont {Dunsworth}}, \bibinfo {author} {\bibfnamefont {E.}~\bibnamefont {Farhi}}, \bibinfo {author} {\bibfnamefont {B.}~\bibnamefont {Foxen}}, \bibinfo {author} {\bibfnamefont {A.}~\bibnamefont {Fowler}}, \bibinfo {author} {\bibfnamefont {C.}~\bibnamefont {Gidney}}, \bibinfo {author} {\bibfnamefont {M.}~\bibnamefont {Giustina}}, \bibinfo {author} {\bibfnamefont {R.}~\bibnamefont {Graff}}, \bibinfo {author} {\bibfnamefont {K.}~\bibnamefont {Guerin}}, \bibinfo {author} {\bibfnamefont {S.}~\bibnamefont {Habegger}}, \bibinfo {author} {\bibfnamefont {M.~P.}\ \bibnamefont {Harrigan}}, \bibinfo {author} {\bibfnamefont {M.~J.}\ \bibnamefont {Hartmann}}, \bibinfo {author} {\bibfnamefont {A.}~\bibnamefont {Ho}}, \bibinfo {author} {\bibfnamefont {M.}~\bibnamefont {Hoffmann}}, \bibinfo {author} {\bibfnamefont {T.}~\bibnamefont {Huang}}, \bibinfo {author} {\bibfnamefont {T.~S.}\ \bibnamefont {Humble}}, \bibinfo {author} {\bibfnamefont {S.~V.}\ \bibnamefont
  {Isakov}}, \bibinfo {author} {\bibfnamefont {E.}~\bibnamefont {Jeffrey}}, \bibinfo {author} {\bibfnamefont {Z.}~\bibnamefont {Jiang}}, \bibinfo {author} {\bibfnamefont {D.}~\bibnamefont {Kafri}}, \bibinfo {author} {\bibfnamefont {K.}~\bibnamefont {Kechedzhi}}, \bibinfo {author} {\bibfnamefont {J.}~\bibnamefont {Kelly}}, \bibinfo {author} {\bibfnamefont {P.~V.}\ \bibnamefont {Klimov}}, \bibinfo {author} {\bibfnamefont {S.}~\bibnamefont {Knysh}}, \bibinfo {author} {\bibfnamefont {A.}~\bibnamefont {Korotkov}}, \bibinfo {author} {\bibfnamefont {F.}~\bibnamefont {Kostritsa}}, \bibinfo {author} {\bibfnamefont {D.}~\bibnamefont {Landhuis}}, \bibinfo {author} {\bibfnamefont {M.}~\bibnamefont {Lindmark}}, \bibinfo {author} {\bibfnamefont {E.}~\bibnamefont {Lucero}}, \bibinfo {author} {\bibfnamefont {D.}~\bibnamefont {Lyakh}}, \bibinfo {author} {\bibfnamefont {S.}~\bibnamefont {Mandr\`{a}}}, \bibinfo {author} {\bibfnamefont {J.~R.}\ \bibnamefont {McClean}}, \bibinfo {author} {\bibfnamefont {M.}~\bibnamefont
  {McEwen}}, \bibinfo {author} {\bibfnamefont {A.}~\bibnamefont {Megrant}}, \bibinfo {author} {\bibfnamefont {X.}~\bibnamefont {Mi}}, \bibinfo {author} {\bibfnamefont {K.}~\bibnamefont {Michielsen}}, \bibinfo {author} {\bibfnamefont {M.}~\bibnamefont {Mohseni}}, \bibinfo {author} {\bibfnamefont {J.}~\bibnamefont {Mutus}}, \bibinfo {author} {\bibfnamefont {O.}~\bibnamefont {Naaman}}, \bibinfo {author} {\bibfnamefont {M.}~\bibnamefont {Neeley}}, \bibinfo {author} {\bibfnamefont {C.}~\bibnamefont {Neill}}, \bibinfo {author} {\bibfnamefont {M.~Y.}\ \bibnamefont {Niu}}, \bibinfo {author} {\bibfnamefont {E.}~\bibnamefont {Ostby}}, \bibinfo {author} {\bibfnamefont {A.}~\bibnamefont {Petukhov}}, \bibinfo {author} {\bibfnamefont {J.~C.}\ \bibnamefont {Platt}}, \bibinfo {author} {\bibfnamefont {C.}~\bibnamefont {Quintana}}, \bibinfo {author} {\bibfnamefont {E.~G.}\ \bibnamefont {Rieffel}}, \bibinfo {author} {\bibfnamefont {P.}~\bibnamefont {Roushan}}, \bibinfo {author} {\bibfnamefont {N.~C.}\ \bibnamefont {Rubin}},
  \bibinfo {author} {\bibfnamefont {D.}~\bibnamefont {Sank}}, \bibinfo {author} {\bibfnamefont {K.~J.}\ \bibnamefont {Satzinger}}, \bibinfo {author} {\bibfnamefont {V.}~\bibnamefont {Smelyanskiy}}, \bibinfo {author} {\bibfnamefont {K.~J.}\ \bibnamefont {Sung}}, \bibinfo {author} {\bibfnamefont {M.~D.}\ \bibnamefont {Trevithick}}, \bibinfo {author} {\bibfnamefont {A.}~\bibnamefont {Vainsencher}}, \bibinfo {author} {\bibfnamefont {B.}~\bibnamefont {Villalonga}}, \bibinfo {author} {\bibfnamefont {T.}~\bibnamefont {White}}, \bibinfo {author} {\bibfnamefont {Z.~J.}\ \bibnamefont {Yao}}, \bibinfo {author} {\bibfnamefont {P.}~\bibnamefont {Yeh}}, \bibinfo {author} {\bibfnamefont {A.}~\bibnamefont {Zalcman}}, \bibinfo {author} {\bibfnamefont {H.}~\bibnamefont {Neven}},\ and\ \bibinfo {author} {\bibfnamefont {J.~M.}\ \bibnamefont {Martinis}},\ }\bibfield  {title} {\bibinfo {title} {Quantum supremacy using a programmable superconducting processor},\ }\href {https://doi.org/10.1038/s41586-019-1666-5} {\bibfield
  {journal} {\bibinfo  {journal} {Nature}\ }\textbf {\bibinfo {volume} {574}},\ \bibinfo {pages} {505} (\bibinfo {year} {2019})}\BibitemShut {NoStop}%
\bibitem [{\citenamefont {Shor}(1997)}]{doi:10.1137/S0097539795293172}%
  \BibitemOpen
  \bibfield  {author} {\bibinfo {author} {\bibfnamefont {P.~W.}\ \bibnamefont {Shor}},\ }\bibfield  {title} {\bibinfo {title} {Polynomial-time algorithms for prime factorization and discrete logarithms on a quantum computer},\ }\href {https://doi.org/10.1137/S0097539795293172} {\bibfield  {journal} {\bibinfo  {journal} {SIAM Journal on Computing}\ }\textbf {\bibinfo {volume} {26}},\ \bibinfo {pages} {1484} (\bibinfo {year} {1997})},\ \Eprint {https://arxiv.org/abs/https://doi.org/10.1137/S0097539795293172} {https://doi.org/10.1137/S0097539795293172} \BibitemShut {NoStop}%
\bibitem [{\citenamefont {Shi}(2005)}]{shi2005quantum}%
  \BibitemOpen
  \bibfield  {author} {\bibinfo {author} {\bibfnamefont {Y.}~\bibnamefont {Shi}},\ }\bibfield  {title} {\bibinfo {title} {Quantum and classical tradeoffs},\ }\href {https://doi.org/10.1016/j.tcs.2005.03.053} {\bibfield  {journal} {\bibinfo  {journal} {Theoretical Computer Science}\ }\textbf {\bibinfo {volume} {344}},\ \bibinfo {pages} {335} (\bibinfo {year} {2005})}\BibitemShut {NoStop}%
\bibitem [{\citenamefont {Demarie}\ \emph {et~al.}(2018)\citenamefont {Demarie}, \citenamefont {Ouyang},\ and\ \citenamefont {Fitzsimons}}]{demarie2018classical}%
  \BibitemOpen
  \bibfield  {author} {\bibinfo {author} {\bibfnamefont {T.~F.}\ \bibnamefont {Demarie}}, \bibinfo {author} {\bibfnamefont {Y.}~\bibnamefont {Ouyang}},\ and\ \bibinfo {author} {\bibfnamefont {J.~F.}\ \bibnamefont {Fitzsimons}},\ }\bibfield  {title} {\bibinfo {title} {Classical verification of quantum circuits containing few basis changes},\ }\bibfield  {journal} {\bibinfo  {journal} {Phys. Rev. A}\ }\textbf {\bibinfo {volume} {97}},\ \href {https://doi.org/10.1103/physreva.97.042319} {10.1103/physreva.97.042319} (\bibinfo {year} {2018})\BibitemShut {NoStop}%
\bibitem [{\citenamefont {Yamakawa}\ and\ \citenamefont {Zhandry}(2022)}]{yamakawa2022verifiable}%
  \BibitemOpen
  \bibfield  {author} {\bibinfo {author} {\bibfnamefont {T.}~\bibnamefont {Yamakawa}}\ and\ \bibinfo {author} {\bibfnamefont {M.}~\bibnamefont {Zhandry}},\ }\bibfield  {title} {\bibinfo {title} {Verifiable quantum advantage without structure},\ }in\ \href {https://doi.org/10.1109/FOCS54457.2022.00014} {\emph {\bibinfo {booktitle} {63rd {IEEE} Annual Symposium on Foundations of Computer Science, {FOCS} 2022, Denver, CO, USA, October 31 - November 3, 2022}}}\ (\bibinfo  {publisher} {{IEEE}, New York},\ \bibinfo {year} {2022})\ pp.\ \bibinfo {pages} {69--74}\BibitemShut {NoStop}%
\bibitem [{\citenamefont {Barak}\ \emph {et~al.}(2012)\citenamefont {Barak}, \citenamefont {Goldreich}, \citenamefont {Impagliazzo}, \citenamefont {Rudich}, \citenamefont {Sahai}, \citenamefont {Vadhan},\ and\ \citenamefont {Yang}}]{10.1145/2160158.2160159}%
  \BibitemOpen
  \bibfield  {author} {\bibinfo {author} {\bibfnamefont {B.}~\bibnamefont {Barak}}, \bibinfo {author} {\bibfnamefont {O.}~\bibnamefont {Goldreich}}, \bibinfo {author} {\bibfnamefont {R.}~\bibnamefont {Impagliazzo}}, \bibinfo {author} {\bibfnamefont {S.}~\bibnamefont {Rudich}}, \bibinfo {author} {\bibfnamefont {A.}~\bibnamefont {Sahai}}, \bibinfo {author} {\bibfnamefont {S.}~\bibnamefont {Vadhan}},\ and\ \bibinfo {author} {\bibfnamefont {K.}~\bibnamefont {Yang}},\ }\bibfield  {title} {\bibinfo {title} {On the (im)possibility of obfuscating programs},\ }\bibfield  {journal} {\bibinfo  {journal} {J. ACM}\ }\textbf {\bibinfo {volume} {59}},\ \href {https://doi.org/10.1145/2160158.2160159} {10.1145/2160158.2160159} (\bibinfo {year} {2012})\BibitemShut {NoStop}%
\bibitem [{\citenamefont {Alagic}\ and\ \citenamefont {Fefferman}(2016)}]{alagic2016quantum}%
  \BibitemOpen
  \bibfield  {author} {\bibinfo {author} {\bibfnamefont {G.}~\bibnamefont {Alagic}}\ and\ \bibinfo {author} {\bibfnamefont {B.}~\bibnamefont {Fefferman}},\ }\href@noop {} {\bibinfo {title} {On quantum obfuscation}} (\bibinfo {year} {2016}),\ \Eprint {https://arxiv.org/abs/1602.01771} {arXiv:1602.01771 [quant-ph]} \BibitemShut {NoStop}%
\bibitem [{\citenamefont {Ananth}\ and\ \citenamefont {Placa}(2021)}]{obfuscation_difficulty_ananth}%
  \BibitemOpen
  \bibfield  {author} {\bibinfo {author} {\bibfnamefont {P.}~\bibnamefont {Ananth}}\ and\ \bibinfo {author} {\bibfnamefont {R.~L.~L.}\ \bibnamefont {Placa}},\ }\bibfield  {title} {\bibinfo {title} {Secure software leasing},\ }in\ \href {https://doi.org/10.1007/978-3-030-77886-6\_17} {\emph {\bibinfo {booktitle} {Advances in Cryptology - {EUROCRYPT} 2021 - 40th Annual International Conference on the Theory and Applications of Cryptographic Techniques, Zagreb, Croatia, October 17-21, 2021, Proceedings, Part {II}}}},\ \bibinfo {series} {Lecture Notes in Computer Science}, Vol.\ \bibinfo {volume} {12697},\ \bibinfo {editor} {edited by\ \bibinfo {editor} {\bibfnamefont {A.}~\bibnamefont {Canteaut}}\ and\ \bibinfo {editor} {\bibfnamefont {F.}~\bibnamefont {Standaert}}}\ (\bibinfo  {publisher} {Springer, Cham, Switzerland},\ \bibinfo {year} {2021})\ pp.\ \bibinfo {pages} {501--530}\BibitemShut {NoStop}%
\bibitem [{\citenamefont {Alagic}\ \emph {et~al.}(2021)\citenamefont {Alagic}, \citenamefont {Brakerski}, \citenamefont {Dulek},\ and\ \citenamefont {Schaffner}}]{obfuscation_difficulty_alagic}%
  \BibitemOpen
  \bibfield  {author} {\bibinfo {author} {\bibfnamefont {G.}~\bibnamefont {Alagic}}, \bibinfo {author} {\bibfnamefont {Z.}~\bibnamefont {Brakerski}}, \bibinfo {author} {\bibfnamefont {Y.}~\bibnamefont {Dulek}},\ and\ \bibinfo {author} {\bibfnamefont {C.}~\bibnamefont {Schaffner}},\ }\bibfield  {title} {\bibinfo {title} {Impossibility of quantum virtual black-box obfuscation of classical circuits},\ }in\ \href {https://doi.org/10.1007/978-3-030-84242-0\_18} {\emph {\bibinfo {booktitle} {Advances in Cryptology - {CRYPTO} 2021 - 41st Annual International Cryptology Conference, {CRYPTO} 2021, Virtual Event, August 16-20, 2021, Proceedings, Part {I}}}},\ \bibinfo {series} {Lecture Notes in Computer Science}, Vol.\ \bibinfo {volume} {12825},\ \bibinfo {editor} {edited by\ \bibinfo {editor} {\bibfnamefont {T.}~\bibnamefont {Malkin}}\ and\ \bibinfo {editor} {\bibfnamefont {C.}~\bibnamefont {Peikert}}}\ (\bibinfo  {publisher} {Springer, Cham, Switzerland},\ \bibinfo {year} {2021})\ pp.\ \bibinfo {pages}
  {497--525}\BibitemShut {NoStop}%
\bibitem [{\citenamefont {Tucci}(2005)}]{KAK}%
  \BibitemOpen
  \bibfield  {author} {\bibinfo {author} {\bibfnamefont {R.~R.}\ \bibnamefont {Tucci}},\ }\href@noop {} {\bibinfo {title} {An introduction to cartan's kak decomposition for qc programmers}} (\bibinfo {year} {2005}),\ \Eprint {https://arxiv.org/abs/quant-ph/0507171} {arXiv:quant-ph/0507171 [quant-ph]} \BibitemShut {NoStop}%
\bibitem [{\citenamefont {Vidal}\ and\ \citenamefont {Dawson}(2004)}]{PhysRevA.69.010301}%
  \BibitemOpen
  \bibfield  {author} {\bibinfo {author} {\bibfnamefont {G.}~\bibnamefont {Vidal}}\ and\ \bibinfo {author} {\bibfnamefont {C.~M.}\ \bibnamefont {Dawson}},\ }\bibfield  {title} {\bibinfo {title} {Universal quantum circuit for two-qubit transformations with three controlled-not gates},\ }\href {https://doi.org/10.1103/PhysRevA.69.010301} {\bibfield  {journal} {\bibinfo  {journal} {Phys. Rev. A}\ }\textbf {\bibinfo {volume} {69}},\ \bibinfo {pages} {010301} (\bibinfo {year} {2004})}\BibitemShut {NoStop}%
\bibitem [{\citenamefont {Shende}\ \emph {et~al.}(2004)\citenamefont {Shende}, \citenamefont {Markov},\ and\ \citenamefont {Bullock}}]{PhysRevA.69.062321}%
  \BibitemOpen
  \bibfield  {author} {\bibinfo {author} {\bibfnamefont {V.~V.}\ \bibnamefont {Shende}}, \bibinfo {author} {\bibfnamefont {I.~L.}\ \bibnamefont {Markov}},\ and\ \bibinfo {author} {\bibfnamefont {S.~S.}\ \bibnamefont {Bullock}},\ }\bibfield  {title} {\bibinfo {title} {Minimal universal two-qubit controlled-not-based circuits},\ }\href {https://doi.org/10.1103/PhysRevA.69.062321} {\bibfield  {journal} {\bibinfo  {journal} {Phys. Rev. A}\ }\textbf {\bibinfo {volume} {69}},\ \bibinfo {pages} {062321} (\bibinfo {year} {2004})}\BibitemShut {NoStop}%
\bibitem [{\citenamefont {Shende}\ \emph {et~al.}(2006)\citenamefont {Shende}, \citenamefont {Bullock},\ and\ \citenamefont {Markov}}]{Shende_2006}%
  \BibitemOpen
  \bibfield  {author} {\bibinfo {author} {\bibfnamefont {V.}~\bibnamefont {Shende}}, \bibinfo {author} {\bibfnamefont {S.}~\bibnamefont {Bullock}},\ and\ \bibinfo {author} {\bibfnamefont {I.}~\bibnamefont {Markov}},\ }\bibfield  {title} {\bibinfo {title} {Synthesis of quantum-logic circuits},\ }\href {https://doi.org/10.1109/tcad.2005.855930} {\bibfield  {journal} {\bibinfo  {journal} {{IEEE} Transactions on Computer-Aided Design of Integrated Circuits and Systems}\ }\textbf {\bibinfo {volume} {25}},\ \bibinfo {pages} {1000} (\bibinfo {year} {2006})}\BibitemShut {NoStop}%
\bibitem [{\citenamefont {Moll}\ \emph {et~al.}(2018)\citenamefont {Moll}, \citenamefont {Barkoutsos}, \citenamefont {Bishop}, \citenamefont {Chow}, \citenamefont {Cross}, \citenamefont {Egger}, \citenamefont {Filipp}, \citenamefont {Fuhrer}, \citenamefont {Gambetta}, \citenamefont {Ganzhorn}, \citenamefont {Kandala}, \citenamefont {Mezzacapo}, \citenamefont {M\"{u}ller}, \citenamefont {Riess}, \citenamefont {Salis}, \citenamefont {Smolin}, \citenamefont {Tavernelli},\ and\ \citenamefont {Temme}}]{Moll_2018}%
  \BibitemOpen
  \bibfield  {author} {\bibinfo {author} {\bibfnamefont {N.}~\bibnamefont {Moll}}, \bibinfo {author} {\bibfnamefont {P.}~\bibnamefont {Barkoutsos}}, \bibinfo {author} {\bibfnamefont {L.~S.}\ \bibnamefont {Bishop}}, \bibinfo {author} {\bibfnamefont {J.~M.}\ \bibnamefont {Chow}}, \bibinfo {author} {\bibfnamefont {A.}~\bibnamefont {Cross}}, \bibinfo {author} {\bibfnamefont {D.~J.}\ \bibnamefont {Egger}}, \bibinfo {author} {\bibfnamefont {S.}~\bibnamefont {Filipp}}, \bibinfo {author} {\bibfnamefont {A.}~\bibnamefont {Fuhrer}}, \bibinfo {author} {\bibfnamefont {J.~M.}\ \bibnamefont {Gambetta}}, \bibinfo {author} {\bibfnamefont {M.}~\bibnamefont {Ganzhorn}}, \bibinfo {author} {\bibfnamefont {A.}~\bibnamefont {Kandala}}, \bibinfo {author} {\bibfnamefont {A.}~\bibnamefont {Mezzacapo}}, \bibinfo {author} {\bibfnamefont {P.}~\bibnamefont {M\"{u}ller}}, \bibinfo {author} {\bibfnamefont {W.}~\bibnamefont {Riess}}, \bibinfo {author} {\bibfnamefont {G.}~\bibnamefont {Salis}}, \bibinfo {author} {\bibfnamefont
  {J.}~\bibnamefont {Smolin}}, \bibinfo {author} {\bibfnamefont {I.}~\bibnamefont {Tavernelli}},\ and\ \bibinfo {author} {\bibfnamefont {K.}~\bibnamefont {Temme}},\ }\bibfield  {title} {\bibinfo {title} {Quantum optimization using variational algorithms on near-term quantum devices},\ }\href {https://doi.org/10.1088/2058-9565/aab822} {\bibfield  {journal} {\bibinfo  {journal} {Quantum Science and Technology}\ }\textbf {\bibinfo {volume} {3}},\ \bibinfo {pages} {030503} (\bibinfo {year} {2018})}\BibitemShut {NoStop}%
\bibitem [{\citenamefont {Suzuki}\ \emph {et~al.}(2021)\citenamefont {Suzuki}, \citenamefont {Kawase}, \citenamefont {Masumura}, \citenamefont {Hiraga}, \citenamefont {Nakadai}, \citenamefont {Chen}, \citenamefont {Nakanishi}, \citenamefont {Mitarai}, \citenamefont {Imai}, \citenamefont {Tamiya}, \citenamefont {Yamamoto}, \citenamefont {Yan}, \citenamefont {Kawakubo}, \citenamefont {Nakagawa}, \citenamefont {Ibe}, \citenamefont {Zhang}, \citenamefont {Yamashita}, \citenamefont {Yoshimura}, \citenamefont {Hayashi},\ and\ \citenamefont {Fujii}}]{Qulacs}%
  \BibitemOpen
  \bibfield  {author} {\bibinfo {author} {\bibfnamefont {Y.}~\bibnamefont {Suzuki}}, \bibinfo {author} {\bibfnamefont {Y.}~\bibnamefont {Kawase}}, \bibinfo {author} {\bibfnamefont {Y.}~\bibnamefont {Masumura}}, \bibinfo {author} {\bibfnamefont {Y.}~\bibnamefont {Hiraga}}, \bibinfo {author} {\bibfnamefont {M.}~\bibnamefont {Nakadai}}, \bibinfo {author} {\bibfnamefont {J.}~\bibnamefont {Chen}}, \bibinfo {author} {\bibfnamefont {K.~M.}\ \bibnamefont {Nakanishi}}, \bibinfo {author} {\bibfnamefont {K.}~\bibnamefont {Mitarai}}, \bibinfo {author} {\bibfnamefont {R.}~\bibnamefont {Imai}}, \bibinfo {author} {\bibfnamefont {S.}~\bibnamefont {Tamiya}}, \bibinfo {author} {\bibfnamefont {T.}~\bibnamefont {Yamamoto}}, \bibinfo {author} {\bibfnamefont {T.}~\bibnamefont {Yan}}, \bibinfo {author} {\bibfnamefont {T.}~\bibnamefont {Kawakubo}}, \bibinfo {author} {\bibfnamefont {Y.~O.}\ \bibnamefont {Nakagawa}}, \bibinfo {author} {\bibfnamefont {Y.}~\bibnamefont {Ibe}}, \bibinfo {author} {\bibfnamefont {Y.}~\bibnamefont {Zhang}},
  \bibinfo {author} {\bibfnamefont {H.}~\bibnamefont {Yamashita}}, \bibinfo {author} {\bibfnamefont {H.}~\bibnamefont {Yoshimura}}, \bibinfo {author} {\bibfnamefont {A.}~\bibnamefont {Hayashi}},\ and\ \bibinfo {author} {\bibfnamefont {K.}~\bibnamefont {Fujii}},\ }\bibfield  {title} {\bibinfo {title} {Qulacs: a fast and versatile quantum circuit simulator for research purpose},\ }\href {https://doi.org/10.22331/q-2021-10-06-559} {\bibfield  {journal} {\bibinfo  {journal} {{Quantum}}\ }\textbf {\bibinfo {volume} {5}},\ \bibinfo {pages} {559} (\bibinfo {year} {2021})}\BibitemShut {NoStop}%
\bibitem [{\citenamefont {Developers}(2022)}]{cirq_developers_2022_7465577}%
  \BibitemOpen
  \bibfield  {author} {\bibinfo {author} {\bibfnamefont {C.}~\bibnamefont {Developers}},\ }\href {https://doi.org/10.5281/zenodo.7465577} {\bibinfo {title} {Cirq}} (\bibinfo {year} {2022}),\ \bibinfo {note} {{See full list of authors on Github: https://github .com/quantumlib/Cirq/graphs/contributors}}\BibitemShut {NoStop}%
\bibitem [{qcu(2023)}]{qcunopt_mori}%
  \BibitemOpen
  \href@noop {} {\bibinfo {title} {qcunopt}},\ \bibinfo {howpublished} {\url{https://github.com/yforest-osaka/qcunopt}} (\bibinfo {year} {2023})\BibitemShut {NoStop}%
\bibitem [{\citenamefont {van~de Wetering}\ and\ \citenamefont {Amy}(2023)}]{vandewetering2023optimising}%
  \BibitemOpen
  \bibfield  {author} {\bibinfo {author} {\bibfnamefont {J.}~\bibnamefont {van~de Wetering}}\ and\ \bibinfo {author} {\bibfnamefont {M.}~\bibnamefont {Amy}},\ }\href@noop {} {\bibinfo {title} {Optimising quantum circuits is generally hard}} (\bibinfo {year} {2023}),\ \Eprint {https://arxiv.org/abs/2310.05958} {arXiv:2310.05958 [quant-ph]} \BibitemShut {NoStop}%
\bibitem [{\citenamefont {Bouland}\ \emph {et~al.}(2019)\citenamefont {Bouland}, \citenamefont {Fefferman}, \citenamefont {Nirkhe},\ and\ \citenamefont {Vazirani}}]{complexity_rcs}%
  \BibitemOpen
  \bibfield  {author} {\bibinfo {author} {\bibfnamefont {A.}~\bibnamefont {Bouland}}, \bibinfo {author} {\bibfnamefont {B.}~\bibnamefont {Fefferman}}, \bibinfo {author} {\bibfnamefont {C.}~\bibnamefont {Nirkhe}},\ and\ \bibinfo {author} {\bibfnamefont {U.}~\bibnamefont {Vazirani}},\ }\bibfield  {title} {\bibinfo {title} {On the complexity and verification of quantum random circuit sampling},\ }\href {https://doi.org/10.1038/s41567-018-0318-2} {\bibfield  {journal} {\bibinfo  {journal} {Nat. Phys.}\ }\textbf {\bibinfo {volume} {15}},\ \bibinfo {pages} {159} (\bibinfo {year} {2019})}\BibitemShut {NoStop}%
\bibitem [{\citenamefont {Boixo}\ \emph {et~al.}(2018)\citenamefont {Boixo}, \citenamefont {Isakov}, \citenamefont {Smelyanskiy}, \citenamefont {Babbush}, \citenamefont {Ding}, \citenamefont {Jiang}, \citenamefont {Bremner}, \citenamefont {Martinis},\ and\ \citenamefont {Neven}}]{boixo2018characterizing}%
  \BibitemOpen
  \bibfield  {author} {\bibinfo {author} {\bibfnamefont {S.}~\bibnamefont {Boixo}}, \bibinfo {author} {\bibfnamefont {S.~V.}\ \bibnamefont {Isakov}}, \bibinfo {author} {\bibfnamefont {V.~N.}\ \bibnamefont {Smelyanskiy}}, \bibinfo {author} {\bibfnamefont {R.}~\bibnamefont {Babbush}}, \bibinfo {author} {\bibfnamefont {N.}~\bibnamefont {Ding}}, \bibinfo {author} {\bibfnamefont {Z.}~\bibnamefont {Jiang}}, \bibinfo {author} {\bibfnamefont {M.~J.}\ \bibnamefont {Bremner}}, \bibinfo {author} {\bibfnamefont {J.~M.}\ \bibnamefont {Martinis}},\ and\ \bibinfo {author} {\bibfnamefont {H.}~\bibnamefont {Neven}},\ }\bibfield  {title} {\bibinfo {title} {Characterizing quantum supremacy in near-term devices},\ }\href {https://doi.org/10.1038/s41567-018-0124-x} {\bibfield  {journal} {\bibinfo  {journal} {Nat. Phys.}\ }\textbf {\bibinfo {volume} {14}},\ \bibinfo {pages} {595} (\bibinfo {year} {2018})}\BibitemShut {NoStop}%
\bibitem [{\citenamefont {Kissinger}\ and\ \citenamefont {van~de Wetering}(2020)}]{Kissinger_2020}%
  \BibitemOpen
  \bibfield  {author} {\bibinfo {author} {\bibfnamefont {A.}~\bibnamefont {Kissinger}}\ and\ \bibinfo {author} {\bibfnamefont {J.}~\bibnamefont {van~de Wetering}},\ }\bibfield  {title} {\bibinfo {title} {{PyZX}: Large scale automated diagrammatic reasoning},\ }\href {https://doi.org/10.4204/eptcs.318.14} {\bibfield  {journal} {\bibinfo  {journal} {Electronic Proceedings in Theoretical Computer Science}\ }\textbf {\bibinfo {volume} {318}},\ \bibinfo {pages} {229} (\bibinfo {year} {2020})}\BibitemShut {NoStop}%
\bibitem [{\citenamefont {Hietala}\ \emph {et~al.}(2023)\citenamefont {Hietala}, \citenamefont {Rand}, \citenamefont {Li}, \citenamefont {Hung}, \citenamefont {Wu},\ and\ \citenamefont {Hicks}}]{10.1145/3604630}%
  \BibitemOpen
  \bibfield  {author} {\bibinfo {author} {\bibfnamefont {K.}~\bibnamefont {Hietala}}, \bibinfo {author} {\bibfnamefont {R.}~\bibnamefont {Rand}}, \bibinfo {author} {\bibfnamefont {L.}~\bibnamefont {Li}}, \bibinfo {author} {\bibfnamefont {S.-H.}\ \bibnamefont {Hung}}, \bibinfo {author} {\bibfnamefont {X.}~\bibnamefont {Wu}},\ and\ \bibinfo {author} {\bibfnamefont {M.}~\bibnamefont {Hicks}},\ }\bibfield  {title} {\bibinfo {title} {A verified optimizer for quantum circuits},\ }\bibfield  {journal} {\bibinfo  {journal} {ACM Trans. Program. Lang. Syst.}\ }\textbf {\bibinfo {volume} {45}},\ \href {https://doi.org/10.1145/3604630} {10.1145/3604630} (\bibinfo {year} {2023})\BibitemShut {NoStop}%
\bibitem [{\citenamefont {Amy}\ and\ \citenamefont {Gheorghiu}(2020)}]{Amy_2020}%
  \BibitemOpen
  \bibfield  {author} {\bibinfo {author} {\bibfnamefont {M.}~\bibnamefont {Amy}}\ and\ \bibinfo {author} {\bibfnamefont {V.}~\bibnamefont {Gheorghiu}},\ }\bibfield  {title} {\bibinfo {title} {staq---a full-stack quantum processing toolkit},\ }\href {https://doi.org/10.1088/2058-9565/ab9359} {\bibfield  {journal} {\bibinfo  {journal} {Quantum Science and Technology}\ }\textbf {\bibinfo {volume} {5}},\ \bibinfo {pages} {034016} (\bibinfo {year} {2020})}\BibitemShut {NoStop}%
\bibitem [{\citenamefont {Smith}\ \emph {et~al.}(2020)\citenamefont {Smith}, \citenamefont {Peterson}, \citenamefont {Skilbeck},\ and\ \citenamefont {Davis}}]{Smith_2020}%
  \BibitemOpen
  \bibfield  {author} {\bibinfo {author} {\bibfnamefont {R.~S.}\ \bibnamefont {Smith}}, \bibinfo {author} {\bibfnamefont {E.~C.}\ \bibnamefont {Peterson}}, \bibinfo {author} {\bibfnamefont {M.~G.}\ \bibnamefont {Skilbeck}},\ and\ \bibinfo {author} {\bibfnamefont {E.~J.}\ \bibnamefont {Davis}},\ }\bibfield  {title} {\bibinfo {title} {An open-source, industrial-strength optimizing compiler for quantum programs},\ }\href {https://doi.org/10.1088/2058-9565/ab9acb} {\bibfield  {journal} {\bibinfo  {journal} {Quantum Science and Technology}\ }\textbf {\bibinfo {volume} {5}},\ \bibinfo {pages} {044001} (\bibinfo {year} {2020})}\BibitemShut {NoStop}%
\bibitem [{\citenamefont {Aaronson}\ and\ \citenamefont {Zhang}(2024)}]{aaronson_peaked}%
  \BibitemOpen
  \bibfield  {author} {\bibinfo {author} {\bibfnamefont {S.}~\bibnamefont {Aaronson}}\ and\ \bibinfo {author} {\bibfnamefont {Y.}~\bibnamefont {Zhang}},\ }\href {https://arxiv.org/abs/2404.14493} {\bibinfo {title} {On verifiable quantum advantage with peaked circuit sampling}} (\bibinfo {year} {2024}),\ \Eprint {https://arxiv.org/abs/2404.14493} {arXiv:2404.14493 [quant-ph]} \BibitemShut {NoStop}%
\bibitem [{\citenamefont {Biamonte}\ \emph {et~al.}(2017)\citenamefont {Biamonte}, \citenamefont {Wittek}, \citenamefont {Pancotti}, \citenamefont {Rebentrost}, \citenamefont {Wiebe},\ and\ \citenamefont {Lloyd}}]{Biamonte_2017}%
  \BibitemOpen
  \bibfield  {author} {\bibinfo {author} {\bibfnamefont {J.}~\bibnamefont {Biamonte}}, \bibinfo {author} {\bibfnamefont {P.}~\bibnamefont {Wittek}}, \bibinfo {author} {\bibfnamefont {N.}~\bibnamefont {Pancotti}}, \bibinfo {author} {\bibfnamefont {P.}~\bibnamefont {Rebentrost}}, \bibinfo {author} {\bibfnamefont {N.}~\bibnamefont {Wiebe}},\ and\ \bibinfo {author} {\bibfnamefont {S.}~\bibnamefont {Lloyd}},\ }\bibfield  {title} {\bibinfo {title} {Quantum machine learning},\ }\href {https://doi.org/10.1038/nature23474} {\bibfield  {journal} {\bibinfo  {journal} {Nature}\ }\textbf {\bibinfo {volume} {549}},\ \bibinfo {pages} {195} (\bibinfo {year} {2017})}\BibitemShut {NoStop}%
\bibitem [{\citenamefont {Nakayama}\ \emph {et~al.}(2023)\citenamefont {Nakayama}, \citenamefont {Mitarai}, \citenamefont {Placidi}, \citenamefont {Sugimoto},\ and\ \citenamefont {Fujii}}]{nakayama2023vqegenerated}%
  \BibitemOpen
  \bibfield  {author} {\bibinfo {author} {\bibfnamefont {A.}~\bibnamefont {Nakayama}}, \bibinfo {author} {\bibfnamefont {K.}~\bibnamefont {Mitarai}}, \bibinfo {author} {\bibfnamefont {L.}~\bibnamefont {Placidi}}, \bibinfo {author} {\bibfnamefont {T.}~\bibnamefont {Sugimoto}},\ and\ \bibinfo {author} {\bibfnamefont {K.}~\bibnamefont {Fujii}},\ }\href@noop {} {\bibinfo {title} {Vqe-generated quantum circuit dataset for machine learning}} (\bibinfo {year} {2023}),\ \Eprint {https://arxiv.org/abs/2302.09751} {arXiv:2302.09751 [quant-ph]} \BibitemShut {NoStop}%
\bibitem [{\citenamefont {Placidi}\ \emph {et~al.}(2023)\citenamefont {Placidi}, \citenamefont {Hataya}, \citenamefont {Mori}, \citenamefont {Aoyama}, \citenamefont {Morisaki}, \citenamefont {Mitarai},\ and\ \citenamefont {Fujii}}]{placidi2023mnisq}%
  \BibitemOpen
  \bibfield  {author} {\bibinfo {author} {\bibfnamefont {L.}~\bibnamefont {Placidi}}, \bibinfo {author} {\bibfnamefont {R.}~\bibnamefont {Hataya}}, \bibinfo {author} {\bibfnamefont {T.}~\bibnamefont {Mori}}, \bibinfo {author} {\bibfnamefont {K.}~\bibnamefont {Aoyama}}, \bibinfo {author} {\bibfnamefont {H.}~\bibnamefont {Morisaki}}, \bibinfo {author} {\bibfnamefont {K.}~\bibnamefont {Mitarai}},\ and\ \bibinfo {author} {\bibfnamefont {K.}~\bibnamefont {Fujii}},\ }\href@noop {} {\bibinfo {title} {Mnisq: A large-scale quantum circuit dataset for machine learning on/for quantum computers in the nisq era}} (\bibinfo {year} {2023}),\ \Eprint {https://arxiv.org/abs/2306.16627} {arXiv:2306.16627 [quant-ph]} \BibitemShut {NoStop}%
\end{thebibliography}%

\end{document}